\documentclass[11pt,onecolumn,letter]{IEEEtran}
\usepackage{graphics,graphicx,color,float}
\usepackage{epsfig}
\usepackage{bm}
\usepackage{mathtools}
\usepackage{amsmath}
\usepackage{amssymb}
\usepackage{amsfonts}
\usepackage{amsthm,hyperref}
\usepackage {times}
\usepackage{setspace}
\usepackage[usenames,dvipsnames]{xcolor}
\usepackage{geometry}

\newcommand{\beq}{\begin{equation}}
\newcommand{\enq}{\end{equation}}
\newcommand{\bel}{\begin{lemma}}
\newcommand{\enl}{\end{lemma}}
\newcommand{\bet}{\begin{theorem}}
\newcommand{\ent}{\end{theorem}}

\newcommand{\tr}{\mathrm{Tr}}

\newcommand{\E}{\mathbb{E}}
\newcommand{\Tr}{\mathrm{Tr}}
\newcommand{\ketbra}[1]{|#1\rangle\langle#1|}

\newcommand{\eps}{\varepsilon}

\newcommand{\id}{\ensuremath{\mathbb{I}}}
 
\def\SU{\mathop{\rm SU}\nolimits}

\newcommand*{\cC}{\mathcal{C}}

\newcommand*{\cH}{\mathcal{H}}

\newcommand*{\cB}{\mathcal{B}}

\newcommand*{\cN}{\mathcal{N}}

\newcommand*{\cX}{\mathcal{X}}

\newcommand*{\cE}{\mathcal{E}}
\newcommand*{\cU}{\mathcal{U}}

\newcommand{\suppress}[1]{}

\newcommand{\defeq}{\ensuremath{ \stackrel{\mathrm{def}}{=} }}
\newcommand{\F}{\mathrm{F}}
\newcommand{\Pur}{\mathrm{P}}
\newcommand {\br} [1] {\ensuremath{ \left( #1 \right) }}

\newcommand{\bra}[1]{\langle #1|}
\newcommand{\ket}[1]{|#1 \rangle}

\mathchardef\mhyphen="2D

\newcommand*{\renyi}{R\'{e}nyi }

\def\sM{\mathsf{M}}
\def\rE{\mathbb{E}}

\makeatletter
\newcommand*{\rom}[1]{\expandafter\@slowromancap\romannumeral #1@}
\makeatother

\mathchardef\mhyphen="2D

\newtheorem{definition}{Definition}

\newtheorem{fact}{Fact}

\newtheorem{theorem}{Theorem}
\newtheorem{lemma}{Lemma}
\newtheorem{corollary}{Corollary}

\begin {document}

\title{Secure communication over fully quantum Gel'fand-Pinsker wiretap channel}
\author{Anurag Anshu, Masahito Hayashi~\IEEEmembership{Fellow, IEEE}, and Naqueeb Ahmad Warsi
\thanks{NW and AA acknowledge the support by the Singapore Ministry of Education and the National Research Foundation, through the Tier 3 Grant Random numbers from quantum processes MOE2012-T3-1-009.
MH was supported in part by JSPS Grants-in-Aid for Scientific Research (A) No.17H01280 and (B) No. 16KT0017 and Kayamori Foundation of Informational Science Advancement.
The Centre for Quantum Technologies is funded
by the Singapore Ministry of Education and the National Research
Foundation as part of the Research Centres of Excellence programme.
The material in this paper was presented in part at the 2018 IEEE International Symposium on Information Theory (ISIT 2018),  
Talisa Hotel in Vail, Colorado, USA, June, 17 -- 22, 2018. \cite{Con}.}
\thanks{
Anurag Anshu is with the Institute for Quantum Computing, University of Waterloo, Waterloo and The Perimeter Institute for Theoretical Physics, Waterloo.
He was with Centre for Quantum Technologies, National University of Singapore, Singapore. 
 (e-mail: aanshu@uwaterloo.ca).
Masahito Hayashi is with the Graduate School of Mathematics, Nagoya University,
Furocho, Chikusaku, Nagoya, 464-860, Japan.
He is also with 
Shenzhen Institute for Quantum Science and Engineering, Southern University of Science and Technology
and the Centre for Quantum Technologies, National University of Singapore, Singapore
(e-mail: masahito@math.nagoya-u.ac.jp).
Naqueeb Ahmad Warsi is with Centre for Quantum Technologies, National University of Singapore, Singapore. (e-mail: warsi.naqueeb@gmail.com).
}}

\markboth{A. Anshu, M. Hayashi, and N. A. Warsi: Secure communication over fully quantum Gel'fand-Pinsker wiretap channel}{}

\maketitle

\maketitle
\begin{abstract}
In this work we study the problem of secure communication over a fully quantum Gel'fand-Pinsker channel. The best known achievability rate for this channel model in the classical case was proven by Goldfeld, Cuff and Permuter in \cite{goldfled-cuff-perumter}. We generalize the result of \cite{goldfled-cuff-perumter}. One key feature of the results obtained in this work is that all the bounds obtained are in terms of error exponent. We obtain our achievability result via the technique of simultaneous pinching. This in turn allows us to show the existence of a simultaneous decoder. Further, to obtain our encoding technique and to prove the security feature of our coding scheme we prove a bivariate classical-quantum channel resolvability lemma and a conditional classical-quantum channel resolvability lemma. As a by product of the achievability result obtained in this work, we also obtain an achievable rate for a fully quantum Gel'fand-Pinsker channel in the absence of Eve. The form of this achievable rate matches with its classical counterpart. The Gel'fand-Pinsker channel model had earlier only been studied for the classical-quantum case and in the case where Alice (the sender) and Bob (the receiver) have shared entanglement between them. 
\end{abstract}
\section{Introduction}
The concept of communication over the wiretap channel was pioneered in the classical case by Wyner \cite{Wyner75}. In this model the wiretapper (Eve) is aware of the encoding strategy used by the transmitter (Alice) to transmit the messages reliably to the legitimate receiver (Bob).  A wiretap channel is classically modeled as a conditional probability distribution $p_{YZ \mid X},$ where $X$ is the channel input supplied by Alice and $(Y,Z)$ are the channel outputs with $Y$ received by Bob and $Z$ received by Eve. The goal here is to maximize the rate of reliable message transmission from Alice to Bob over this channel, such that Eve gets to know as little information as possible about the transmitted message.

This problem of secure communication over noisy wiretap channel was extended to the quantum domain in \cite{cai-winter-yeung, devetak-2005}. In the quantum case, the wiretap channel is modeled as a CPTP (completely positive and trace preserving) map $\cN_{A \to BE},$ where $A$ is the input register supplied by Alice and $B$ and $E$ represent Bob's and Eve's respective shares of the channel output. The quantum wiretap channel model has also been well studied in the one-shot scenario, see for example \cite{renes-renner-2011, Hayashi-wiretap, Wil17a, rsw17}.  

Recently, there has been an interest in studying the classical wiretap channel with \emph{states}.  A classical wiretap channel with states is modeled as $p_{YZ \mid XS}.$ Similar to the wiretap channel as discussed above, it produces two outputs $(Y,Z)$ with $Y$ received by Bob and $Z$ received by Eve. However, unlike the normal wiretap channel, in this case the channel takes two inputs $X$ (supplied by Alice) and a random parameter $S$. This random parameter $S$ is used to represent the channel state and is not controlled by the transmitter.  A key motivation for studying this channel model is that it captures the scenario of communication both in the presence of a jammer and an eavesdropper.  Further, this channel also models the scenario of  a memory with stuck-at faults (for more details, please see \cite {Frederic10}). In \cite{Chen-Vinck-2008}, Chen and Vinck considered the problem of communication over this channel model in the presence of an eavesdropper. 
They combined the coding strategy for the normal wiretap channel along with the coding strategy for the Gel'fand-Pinsker (GP) channel and obtained a lower bound on the secrecy capacity. In \cite{Chia-Gammal-2012}, Chia and El-Gamal further advanced the theory of communication over this channel model by proposing a more sophisticated coding technique in the case when the full channel side information is causally available at both the encoder and the decoder. Even though their coding strategy is restricted to utilize the state information in a causal manner, the authors show that their technique allows to achieve a better transmission rate as compared to the one obtained in \cite{Chen-Vinck-2008}.  The GP channel model in the absence of Eve was first introduced by Gel'fand and Pinsker in their seminal work \cite{GelfandP80}. In this model there are two parties (Alice and Bob) and it takes two inputs $X$ (supplied by Alice) and a random parameter $S$. However, unlike the GP-wiretap channel model this channel only produces one output $Y$ received by Bob. 

In \cite{goldfled-cuff-perumter} Goldfeld, Cuff and Permuter revisit this communication problem when the channel state side information is causally available at the encoder. The authors motivate this model by noting that having information about the extra randomness $S$ (the channel state parameter) of the channel may help in secure transmission. They employ an encoding technique based on the \emph{superposition} coding scheme \cite{GamalK12} and obtain the best known lower bound on the secrecy capacity of the GP channel model and also recover the results of \cite{Chen-Vinck-2008} and \cite{Chia-Gammal-2012} as a special case. The converse result for this problem is not known except for some special cases (which do not seem to have natural interpretation in the quantum case). To obtain their results, the authors prove what they call a superposition covering lemma.
Although the papers \cite{goldfled-cuff-perumter,Chen-Vinck-2008,Chia-Gammal-2012} 
call the approximation of the output distribution
a covering lemma, 
this type of approximation was studied with the name of channel resovability in the earlier papers \cite{han-verdu-93, hayashi-resolvability, bloch-laneman-resolvability, Hayashi-wiretap}.

\begin{figure}
\centering
\includegraphics[scale=0.4]{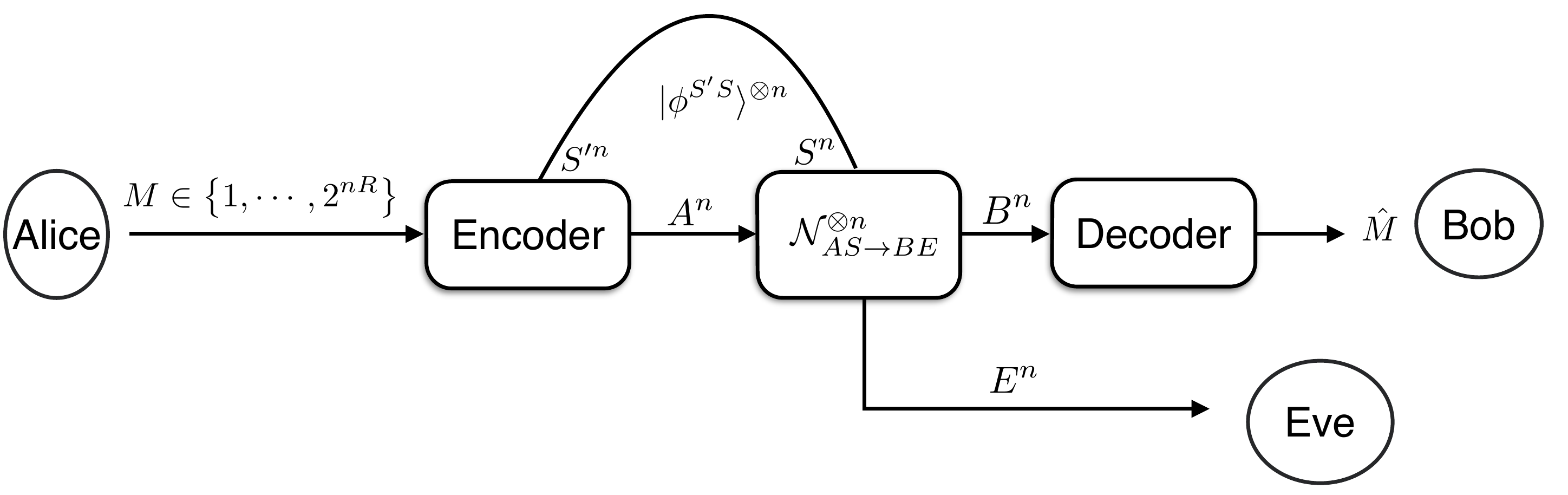}
\caption{{A general block diagram for communication over $n$ independent uses of the Gel'fand-Pinsker wiretap channel. In this model the encoder shares $n$ copies of the entangled state $\ket{\phi^{S'S}}$ with the channel itself, where the register $S'^n$ is held by the encoder and the register $S^n$ is held by the channel. }}
\label{fig:tomcoverellipse}
\end{figure}

We study the problem of secure communication over the fully quantum Gel'fand-Pinsker wiretap channel and provide an exact quantum generalization of the results obtained in \cite{goldfled-cuff-perumter}. Fig \ref{fig:tomcoverellipse} models this communication scheme.
To derive the quantum generalization of the coding technique in \cite{goldfled-cuff-perumter} we prove a generalization of classical-quantum channel resolvability lemma \cite{Hayashi-wiretap}.
This lemma is the quantum analogue of the \cite[Lemma 7]{goldfled-cuff-perumter}.  
To prove the secrecy property of our coding technique we prove a conditional classical-quantum channel resolvability lemma. For the task of designing the decoding POVMs for our protocol we use the technique of \emph{simultaneous pinching} (see \cite{Hayashi-02} for details on the concept of pinching). Using this technique of simultaneous pinching we exhibit the existence  of a \emph{simultaneous decoder} in the single-shot case.   One key feature of the single-shot bounds derived in this manuscript is that they are in terms of \emph{error exponent}. The problem of reliable communication with no security constraint and in the presence of entanglement assistance was first studied in \cite {Frederic10}.

%
The result obtained in this manuscript allows us to recover the previous known results for classical message transmission over point-to-point quantum channels \cite{holevo-channel, SchumacherBWM97} and the quantum wiretap channel (in the absence of the channel state) \cite{cai-winter-yeung, devetak-2005}. Further, our result also implies an achievable rate for classical communication over fully quantum Gel'fand-Pinsker channel (in the absence of Eve). The form of our achievable rate for this problem is exactly similar to that obtained in \cite{GelfandP80}. We note here that the fully quantum Gel'fand-Pinsker channel has been studied in \cite{Frederic10, AnshuJW17CC} only in the case when Alice and Bob \emph{share entanglement} and in \cite{Wc-2017} for classical-quantum channels. Our work is the first work to study this model in the \emph{absence} of entanglement assistance between Alice and Bob. We discuss these results in Corollary \ref{cor:cons}.

\section{Preliminaries}
\label{sec:prelim}

Consider a finite dimensional Hilbert space $\cH$ endowed with an inner product $\langle \cdot, \cdot \rangle$ (in this paper, we only consider finite dimensional Hilbert-spaces). The $\ell_1$ norm of an operator $X$ on $\cH$ is $\| X\|_1:=\Tr\sqrt{X^{\dagger}X}$ and $\ell_2$ norm is $\| X\|_2:=\sqrt{\Tr XX^{\dagger}}$. A quantum state (or a density matrix or a state) is a positive semi-definite matrix on $\cH$ with trace equal to $1$. It is called {\em pure} if and only if its rank is $1$. A sub-normalized state is a positive semi-definite matrix on $\cH$ with trace less than or equal to $1$. Let $\ket{\psi}$ be a unit vector on $\cH$, that is $\langle \psi,\psi \rangle=1$.  With some abuse of notation, we use $\psi$ to represent the state and also the density matrix $\ketbra{\psi}$, associated with $\ket{\psi}$. Given a quantum state $\rho$ on $\cH$, {\em support of $\rho$}, called $\text{supp}(\rho)$ is the subspace of $\cH$ spanned by all eigen-vectors of $\rho$ with non-zero eigenvalues.
 
A {\em quantum register} $A$ is associated with some Hilbert space $\cH_A$. Define $|A| := \dim(\cH_A)$. Let $\mathcal{L}(A)$ represent the set of all linear operators on $\cH_A$. Let $\mathcal{P}(A)$ represent the set of all positive semidefinite operators on $\cH_A$. We denote by $\mathcal{D}(A)$, the set of quantum states on the Hilbert space $\cH_A$. State $\rho$ with subscript $A$ indicates $\rho_A \in \mathcal{D}(A)$. If two registers $A,B$ are associated with the same Hilbert space, we shall represent the relation by $A\equiv B$.  Composition of two registers $A$ and $B$, denoted $AB$, is associated with Hilbert space $\cH_A \otimes \cH_B$.  For two quantum states $\rho\in \mathcal{D}(A)$ and $\sigma\in \mathcal{D}(B)$, $\rho\otimes\sigma \in \mathcal{D}(AB)$ represents the tensor product (Kronecker product) of $\rho$ and $\sigma$. The identity operator on $\cH_A$ (and associated register $A$) is denoted $\id_A.$

Let $\rho_{AB} \in \mathcal{D}(AB)$. We define
\[ \rho_{B} := \Tr_{A}\rho_{AB}
:= \sum_i (\bra{i} \otimes \id_{B})
\rho_{AB} (\ket{i} \otimes \id_{B}) , \]
where $\{\ket{i}\}_i$ is an orthonormal basis for the Hilbert space $\cH_A$.
The state $\rho_B\in \mathcal{D}(B)$ is referred to as the marginal state of $\rho_{AB}$. Unless otherwise stated, a missing register from subscript in a state will represent partial trace over that register. Given a $\rho_A\in\mathcal{D}(A)$, a {\em purification} of $\rho_A$ is a pure state $\rho_{AB}\in \mathcal{D}(AB)$ such that $\Tr_{B}{\rho_{AB}}=\rho_A$. Purification of a quantum state is not unique.
A quantum {map} $\cE: \mathcal{L}(A)\rightarrow \mathcal{L}(B)$ is a completely positive and trace preserving (CPTP) linear map (mapping states in $\mathcal{D}(A)$ to states in $\mathcal{D}(B)$). 
A {\em unitary} operator $U_A:\cH_A \rightarrow \cH_A$ is such that $U_A^{\dagger}U_A = U_A U_A^{\dagger} = \id_A$. An {\em isometry}  $V:\cH_A \rightarrow \cH_B$ is such that $V^{\dagger}V = \id_A$ and $VV^{\dagger} = \id_B$. 

Our model is given as the following pair.
One is a CPTP map $\cN_{AS \to BE}$ from the joint system $(A, S)$ to the joint system $(B,E)$,
where $A$ is the input system, $S$ is the channel internal system,
$B$ is the legitimate receiver (Bob)'s system, and 
$E$ is the wiretapper (Eve)'s system.
The other is an entangled state $|\phi^{S'S}\rangle$ across 
the channel internal system $S$ and the system $S'$ of side information available to the transmitter (Alice).
Using the information in $S'$, Alice can choose the encoder dependently of the channel internal system $S$.
That is, the pair of a CPTP map $\cN_{AS \to BE}$ and an entangled state $|\phi^{S'S}\rangle$ 
gives our model.

\begin{definition}
We shall consider the following information theoretic quantities. 
\begin{enumerate}
\item {\bf Fidelity} (\cite{Josza94}, see also \cite{uhlmann76}) For $\rho_A,\sigma_A \in \mathcal{D}(A)$, $$\F(\rho_A,\sigma_A)\defeq\|\sqrt{\rho_A}\sqrt{\sigma_A}\|_1.$$ 
\item {\bf Purified distance} (\cite{GilchristLN05}) For $\rho_A,\sigma_A \in \mathcal{D}(A)$, $$\Pur(\rho_A,\sigma_A) = \sqrt{1-\F^2(\rho_A,\sigma_A)}.$$ This is different from the Hellinger distance which is defined as $\sqrt{1-\F(\rho_A,\sigma_A}).$

\item {\bf{Sandwiched \renyi relative entropies }}(\cite{wilde-winter-Yang, lenert-Dupuis-Szehr-Tomamichel}) Let $\rho,\sigma \in \mathcal{D}(A)$ and let $\alpha > 0$ we define the following two kinds of \renyi relative entropies:

\begin{align*}
\underline{\mathrm{D}}_{1+\alpha}(\rho\|\sigma)
&:= \frac{1}{\alpha}\log 
\Tr( \sigma^{-\frac{\alpha}{2(1+\alpha)}}\rho \sigma^{-\frac{\alpha}{2(1+\alpha)}})^{1+\alpha}.
\end{align*}

\item {\bf \renyi mutual information and \renyi conditional mutual information (\cite{hayashi-marco-2017})} Let 
 \begin{align*}
 \rho_{UVB} :=& \sum_{u,v}p_{UV}(u,v)\ket{u}\bra{u}_U\otimes\ket{v}\bra{v}_V\otimes \rho_{B \mid u,v};\\
 \rho_{V-U-B} := &\sum_{u}p_{U}(u)\ket{u}\bra{u}_U\otimes\rho_{V \mid u}\otimes \rho_{B \mid u},
 \end{align*}
where in the above $\rho_{V \mid u}$ and $\rho_{B \mid u}$ are appropriate marginals with respect to the state $\rho_{UVB}.$
We define the \renyi mutual information
\begin{align*}
\underline{I}_{1+\alpha}(UV; B)_{\rho_{UVB}|\rho_{UV}}
:=
\min_{\sigma_{B}}\underline{\mathrm{{D}}}_{1+\alpha}(\rho_{UVB}\|\rho_{UV}
\otimes \sigma_{B}),
\end{align*}
where $\sigma_{B}$ is an arbitrary state on $\cH_S$.
Also, we define the \renyi conditional mutual information
\begin{align}
\underline{I}_{1+\alpha}(V;B|U )_{\rho_{UVB}|\rho_{UV}}
:=
\min_{\sigma_{UVB}}\underline{\mathrm{D}}_{1+\alpha}(\rho_{UVB}\|\sigma_{UVB}),
\label{Min11}
\end{align}
where $\sigma_{UVB}$ is given 
with an arbitrary state $\sigma_{B \mid u}$ as 
\begin{align*}
\sigma_{UVB}
= \sum_{u}p_U(u) |u\rangle \langle u| \otimes \rho_{V \mid u}\otimes \sigma_{B \mid u}.
\end{align*}

\end{enumerate}
\label{def:infquant}
\end{definition}	
We will use the following facts. 

\begin{fact}[Minimum achieving state,~\cite{hayash-marco-2016}]
\label{fact:minimum}
For $1+\alpha\ge \frac{1}{2} $, the minimum in \eqref{Min11} is uniquely attained when
$\sigma_{B|u}$ satisfies
\begin{align}
\label{condis}
\sigma_{B|u}=
\frac{
\Tr_A [ 
(\rho_{V \mid u} \otimes \rho_{B \mid u})^{-\frac{\alpha}{2(1+\alpha)}}
\rho_{VB \mid u}
(\rho_{V|u} \otimes \rho_{B \mid u})^{-\frac{\alpha}{2(1+\alpha)}}
]}{
\Tr [ 
(\rho_{V  \mid u} \otimes \rho_{B \mid u})^{-\frac{\alpha}{2(1+\alpha)}}
\rho_{VB \mid u}
(\rho_{V  \mid u } \otimes \rho_{B \mid u})^{-\frac{\alpha}{2(1+\alpha)}}
]} .
\end{align}
Lemma 5 of \cite{hayash-marco-2016} showed the above inequality without the classical system $U$.
Since $U$ is a classical system, we can apply Lemma 5 of \cite{hayash-marco-2016} to the state $\rho_{VB|u}$ for each element $u$, which implies \eqref{condis}.
\end{fact}

\begin{fact}[Triangle inequality for purified distance,~\cite{Tomamichel12}]
\label{fact:trianglepurified}
For states $\rho_A, \sigma_A, \tau_A\in \mathcal{D}(A)$,
$$\Pur(\rho_A,\sigma_A) \leq \Pur(\rho_A,\tau_A)  + \Pur(\tau_A,\sigma_A) ,$$ 
which implies that
$$\Pur(\rho_A,\sigma_A)^2 \leq (\Pur(\rho_A,\tau_A)  + \Pur(\tau_A,\sigma_A) )^2
\leq 2 (\Pur(\rho_A,\tau_A)^2  + \Pur(\tau_A,\sigma_A) ^2).$$ 
\end{fact}
\suppress{
\begin{fact}[\cite{stinespring55}](\textbf{Stinespring representation})\label{stinespring}
Let $\E(\cdot): \mathcal{L}(A)\rightarrow \mathcal{L}(B)$ be a quantum operation. There exists a register $C$ and an unitary $U\in \mathcal{U}(ABC)$ such that $\E(\omega)=\Tr_{A,C}\br{U (\omega  \otimes \ketbra{0}^{B,C}) U^{\dagger}}$. Stinespring representation for a channel is not unique. 
\end{fact}
}
\begin{fact}[Monotonicity under quantum operations, \cite{barnum96},\cite{lindblad75}]
	\label{fact:monotonequantumoperation}
For quantum states $\rho$, $\sigma$ and quantum operation $\cE(\cdot):\mathcal{L}(A)\rightarrow \mathcal{L}(B)$, it holds that
\begin{align*}
\mathrm{\underline{D}}_{1+\alpha}(\cE(\rho)\| \cE(\sigma)) \leq \mathrm{D}_{1+\alpha}(\rho\|\sigma) 
\quad \mbox{and} \quad \Pur(\cE(\rho),\cE(\sigma)) \leq \Pur(\rho,\sigma) 
\end{align*}
\end{fact}

\begin{fact}[Uhlmann's Theorem, \cite{uhlmann76}]
\label{uhlmann}
Let $\rho_A,\sigma_A\in \mathcal{D}(A)$. Let $\rho_{AB}\in \mathcal{D}(AB)$ be a purification of $\rho_A$ and $\ket{\sigma}_{AC}\in\mathcal{D}(AC)$ be a purification of $\sigma_A$. There exists an isometry $V: C \rightarrow B$ such that,
 $$\F(\ketbra{\theta}_{AB}, \ketbra{\rho}_{AB}) = \F(\rho_A,\sigma_A) ,$$
 where $\ket{\theta}_{AB} = (\mathbb{I}_A \otimes V) \ket{\sigma}_{AC}$.
\end{fact}

\begin{fact}
\label{pinsker}
For quantum states $\rho_A,\sigma_A\in\mathcal{D}(A)$, 
$$\F^2(\rho,\sigma) \geq 2^{-\mathrm{\underline{D}}_{1+\alpha}\left(\rho \| \sigma\right)}. $$
The fact follows from \cite [Lemma 5]{Jain:2003a}, see also \cite[Corollary 4.3,]{tomamichel16} and from the monotonicity of sandwiched \renyi relative entropy. 
\end{fact}	

\begin{fact}
\label{purl1}
Let $\rho$ and $\sigma$ be two quantum states. We have the following relation:
\beq
\Pur\left( \rho, \sigma\right) \leq \sqrt{2 \|\rho-\sigma\|_1}.
\enq
\end{fact}

\begin{fact}[~\cite{AnshuJW17CC}]
\label{err}
Let $\rho$ and $\sigma$ be quantum states. Then, for every let $0< \Lambda< \mathbb{I}$ be an operator,
\beq
|\sqrt{\tr\left[\Lambda\rho\right]}-\sqrt{\tr\left[\Lambda\sigma\right]}| \leq \Pur(\rho,\sigma). \nonumber
\enq
\end{fact}

\begin{fact}[Hayashi-Nagaoka inequality, \cite{HyashiN03}]
\label{haynag}
Let $0<S<\id,T$ be positive semi-definite operators. Then 
$$\mathbb{I} - (S+T)^{-\frac{1}{2}}S(S+T)^{-\frac{1}{2}}\leq 2(\mathbb{I}-S) + 4T.$$

\end{fact}

\begin{fact}[Hayashi, \cite{Hayashi-02}]
\label{pinchs}
Let $\rho$ and $\sigma$ be two quantum states. Further, let $\cE$ be the pinching operation with respect to the basis of $\sigma$. Then, 
\beq
\rho \leq v \cE(\rho), \nonumber
\enq
where $v$ represents the distinct number of eigenvalues of $\sigma$ and is sometimes also called as the pinching constant.
\end{fact}

\begin{fact}(Jensen's inequality)
\label{jenssen}
Let $f : \cX \to \mathbb{R}$ be a concave function. Then, $\mathbb{E}_{X}[f(X)] \leq f(\mathbb{E}[X]).$
\end{fact}

\section{Main result}
Before giving our main result we first give the following definitions:
\begin{definition} (Encoding, Decoding,  Error, Secrecy)
\label{edes}
An $(n,M_n, \eps_n,\delta_n)$ secrecy code for communication over a quantum Gel'fand-Pinsker wiretap channel $\cN^{\otimes n}_{AS \to BE}$ with channel state side information available at the encoder (i,e, when the sender shares an entangled state $\ket{\phi^{S'S}}^{\otimes n}$ with the channel itself) consists of  
\begin{itemize}
\item an encoding operation (for Alice) $\cE: M S'^n \to A^n,$ where $S'^n \equiv S^n$ and $|M|= M_n,$ such that 
$$\Pur\left( \rho_{ME^n} ,  \tr_{E^n}[\rho_{ME^n}] \otimes  \tr_{M}[\rho_{ME^n}]\right) \leq \delta_n,$$ where $\rho_{ME^n}:= \frac{1}{M_n} \sum_{m \in [1:M_n]} \ket{m}\bra{m}_M \otimes \mathcal{N}^{\otimes n}_{AS \to BE} \left(\cE \left(m,S'^n\right),S^n\right)$ and $\Pur\left(\cdot, \cdot\right)$ is the purified distance.
\item a decoding operation (for Bob) $\mathcal{D}: B^n \to \hat{M}_n,$ with $\hat{M}_n \equiv M_n.$ such that
$$\Pr \left\{ M \neq \hat{M}\right\} \leq \eps_n$$

\end{itemize}
\end{definition}
\begin{definition}
A rate $R$ is said to be achievable if there exists a sequence of $(n,M_n, \eps_n,\delta_n)$- codes such that 
\begin{align*}
 \liminf_{n \to \infty}\frac{1}{n} \log M_n & \geq R; \\
 \limsup_{n \to \infty } \eps_n &\to 0; \\
 \lim _{n \to \infty} \delta_n& \to 0.
\end{align*} 
The supremum of all the achievable rates is called the secrecy capacity of the Gel'fand-Pinsker channel.  
\end{definition}

The following theorem is one of our main result. It can be considered as the quantum generalization of the achievability result in \cite[Equations $22$ and $24$]{goldfled-cuff-perumter}.
\begin{theorem}
\label{achievabilityasy}
Let $\cN_{AS \to BE}$ be a quantum Gel'fand-Pinsker wiretap channel. Further, let $\ket{\phi}_{S'S}$ be the shared entanglement between the sender and the channel. 
We choose a joint distribution $p_{UV}$ and conditional states $\{\rho_{AS \mid u,v} \}_{u,v}$ 
such that $\tr_{UVA}{\rho_{UVAS}}= \tr_{S'}\phi_{S'S}$, where 
$\rho_{UVAS}:=\sum_{(u,v)}p_{UV}(u,v) \ket{u}\bra{u} _U\otimes \ket{v}\bra{v}_V \otimes \rho_{AS \mid u,v}$. 
Then,
a rate $R$ is achievable if 
\begin{align}
R \leq  R_a(\rho_{UVAS}):= \min \left\{I [V;B \mid U]- I[V;E \mid U], I[UV;B] - I[UV;S],
I[UV;B] - I[U;S]- I[V;E \mid U]
\right\},
\label{NHFG}
\end{align}
where the information theoretic quantities above are calculated with respect to the state 
$\rho_{UVAS}$.
\end{theorem} 

We denote the set of 
$\rho_{UVAS}$ to satisfy the condition given in Theorem \ref{achievabilityasy}
by ${\cal S}_1$.
Then, the rate $$\max_{\rho_{UVAS} \in {\cal S}_1}R_a(\rho_{UVAS}) $$ is achievable.
To simplify this rate, we introduce the set ${\cal S}_2:=
\{\rho_{UVAS} \in {\cal S}_1| I[U;B] -I(U;S)\geq 0 \}$.
Then, we have the following lemma.
\begin{lemma}\label{LA}
\begin{align}
\max_{\rho_{UVAS} \in {\cal S}_1} R_a(\rho_{UVAS})
=
\max_{\rho_{UVAS} \in {\cal S}_2} R_{\mbox{alt}}(\rho_{UVAS}),
\end{align}
where
\begin{align}
\label{altttt}
R_{\mbox{alt}}(\rho_{UVAS}):=
 \min \left\{I [V;B \mid U]- I[V;E \mid U], I[UV;B] - I[UV;S]\right\}.
\end{align}
\end{lemma}
Therefore, 
Theorem \ref{achievabilityasy} guarantees that the rate
$\max_{\rho_{UVAS} \in {\cal S}_2} R_{\mbox{alt}}(\rho_{UVAS})$ is also achievable.
The proof of this lemma follows exactly similar to that given in \cite[Appendix A]{goldfled-cuff-perumter}. However, we repeat the same proof in the Appendix just for completeness.

To achieve the rate given in \eqref{NHFG},
we employ superposition coding, in which we randomly choose $U$ and we make an encoder with respect to $V$ conditioned on $U$.
Here, we elaborate upon the roles of $U$ and $V$. In Gel'fand-Pinsker wiretap channel, the register $S$ some correlation with 
the systems $E$ and $B$, which makes our analysis difficult. We convert these correlations to the correlation between
the register $S$ and the message. Therefore, we need three types of evaluations. The first one is the evaluation of the correlation between the register $S$ and the message. The second one is the evaluation of the decoding error probability 
with ignoring the correlation between the register $S$ and the receiver $B$.
It can be evaluated as a correlation between $U, V$ and $S$. The third one is the evaluation of the information leakage
while ignoring the correlation between the register $S$ and the eavesdropper $E$.
It can be evaluated as the correlation between $V$ and $E$ conditioned with $U$.

To realize the third type of evaluation, we need a scramble variable related to $V$ with the rate $R_1$.
This type of analysis requires the condition
\begin{align}
\label{Arb5}
R_1 > I [V;E \mid U].
\end{align}
In contrast, to realize the second type of analysis, we need another scramble variable related to 
$U$ with the rate $r$ 
as well as the scramble variable related to $V$ with the rate $R_1$.
This type of analysis requires the conditions
\begin{align}
\label{Arb4}
r & > I[U;S] ;\\
\label{Arb3}
R_1+ r & > I[UV;S].
\end{align}
In addition, 
the first type of analysis requires the condition
for the coding rate $R$;
\begin{align}
\label{Arb1}
R+R_1+r & < I[UV;B];\\
\label{Arb2}
R+R_1 & < I[V;B \mid U].
\end{align}  
As explained in the final part of our proof,
combining the conditions \eqref{Arb5} -- \eqref{Arb2}, we can show that
the rate given in \eqref{NHFG} is achievable.

An important consequence of our achievability result is the following corollary:  

\begin{corollary}
\label{cor:cons}
\begin{itemize}
\item[$(a)$] (Communication over point-to-point channel, \cite{holevo-channel, SchumacherBWM97}) Let $\cN_{A \to B}$ be a quantum channel. Further, let $\chi(\cN):= \max_{\rho} I [X;B],$ where the maximization is over the states of the following form: $\sum_{x}p_{X}(x) \ket{x}\bra{x}_{X}\otimes \cN_{A \to B}(\rho_{A \mid x)}.$  Then every rate $R$ satisfying the following constraint
 $$R \leq \lim_{k \to \infty}\frac{1}{k} \chi(\cN^{\otimes k})$$ is achievable.
\item[ {$(b)$} ] (Communication over point-to-point wiretap channel, \cite{cai-winter-yeung, devetak-2005}) Let $\cN_{A \to BE}$ be a quantum wiretap channel. Further, let $P(\cN):=\max_{\rho}\left(I[X;B] - I[X;E]\right),$ where the maximization is over the states of the following form: $\sum_{x}p_{X}(x) \ket{x}\bra{x}_{X}\otimes \cN_{A \to BE}(\rho_{A \mid x)}.$ Then every rate $R$ satisfying the following constraint $$R \leq \lim_{k \to \infty}\frac{1}{k} P(\cN^{\otimes k})$$ is achievable for the wiretap channel $\cN_{A \to BE}$.  
\item [$(c)$] (Entanglement unassisted communication over  Gel'fand-Pinsker quantum channel) Let $\cN_{AS \to B}$ be a Gel'fand-Pinsker quantum channel.  Further, let $\chi(\cN)_{\mbox{GP}}:= \max_{\rho} \left(I[U;B] - I[U;S]\right),$ where the maximization is over the states of the following form: $\rho_{UAS}=\sum_{u}p_{UV}(u,v) \ket{u}\bra{u} _U \otimes \rho_{AS \mid u}$ such that $\tr_{UA}{\rho_{UAS}}= \tr_{S'}\phi_{S'S}.$ Then any rate $R$ satisfying the following constraint 
$$
R \leq \lim_{k \to \infty}\frac{1}{k} \chi_{\mbox{GP}}(\cN^{\otimes k})
$$
is achievable for the channel $\cN_{AS \to B}$.
\end{itemize}
\end{corollary}

\begin{proof}
\begin{description}
\item[$(a)$]  The proof follows by setting $U=\emptyset, V = X, S=\emptyset, E= \emptyset$ in \eqref{altttt} and then using the coding strategy in the proof of Theorem \ref{achievabilityasy} for $\cN_{A \to B}^{\otimes k}.$ 
\item[$(b)$]   The proof follows by setting $U= \emptyset, V = X, S=\emptyset$ in  \eqref{altttt} and then using the coding strategy in the proof of Theorem \ref{achievabilityasy} for $\cN_{A \to BE}^{\otimes k}.$
\item [$(c)$] The proof follows by setting $ V = X, E=\emptyset$ in \eqref{altttt} and then using the coding strategy in the proof of Theorem \ref{achievabilityasy} for $\cN_{AS \to B}^{\otimes k}.$
\end{description}
This completes the proof.
\end{proof}
Before giving the proof of Theorem \ref{achievabilityasy} we first study the single-shot version of the task mentioned in Fig \ref{fig:tomcoverellipse}. For the single-shot case we obtain an error exponent like bound on the decoding error probability and the secrecy criterion. 

\section{Code construction in single-shot form}\label{SS4}
In this section, we give the construction of our code in the single-shot form,
and evaluate its performance.
Let $\cN_{AS \to BE}$ be a quantum Gel'fand-Pinsker wiretap channel. Further, let $\ket{\phi}_{S'S}$ be the shared entanglement between the sender and the channel. 

Let $\rho_{UVAS}$ be as defined in Theorem \ref{achievabilityasy} and define the following states: 
\begin{align}
\rho_{UVBE} &:= \cN_{AS \to BE}(\rho_{UVAS});\\
\label{savegragestate}
\rho_{B} &:= \tr_{UVE} \cN_{AS \to BE}(\rho_{UVAS});\\
\label{saverageu}
\rho_{B\mid u} &:=  \sum_{v}p_{V \mid U}(v \mid u) \rho_{B \mid u,v}.
\end{align}
\noindent{\bf{The codebook:}}
We choose real numbers $R,R_1,r >0$.
Let $U(1), \cdots$, $U(2^{r})$ be drawn independently according to $p_U.$ Further, for every $i \in [1:2^{r}]$ and for every message $m \in [1:2^{R}],$ generate $V(m,i,1), \cdots,$ $V(m,i,2^{R_1})$ independently, where for every $j \in [1:2^{R_1}],$ $V(m,i,j) \sim p_{V \mid U(i)}.$ The distribution $p_{V \mid U (i)}$ is with respect to the conditional distribution of the joint distribution $p_{UV}.$ In what follows we will use the notation $\mathcal{C}_{U}:=  \left\{U(1), U(2), \cdots, U(2^{r})\right\}$ and $\cC_{m,i}:= \left\{V(m,i,1), \cdots, V(m,i,2^{R_1})\right\}$.  Both $\mathcal{C}_{U}$ and $\cC_{m,i}$ for all $(m,i) \in [1:2^{r}] \times [1:2^{R}],$ are known to all the parties involved, i.e., Alice, Bob and Eve.   We will use the notation $\cC_m:= \cup_{i} \cC_{m,i}.$
\vspace{3mm}

\noindent {\bf Remark: } In the above $R$ stands for the final rate of communication. Our encoding scheme mentioned below is a multi-level coding scheme which has the dual feature of being a good wiretap channel code along with being a good code for the Gel'fand-Pinsker channel. Intuitively, the coding scheme should be such that it should be able to keep the messages secure from Eve. Further, since Bob does not have any information about $S$ therefore the encoding scheme should be such that it should somehow help Bob in decoding. These two features about our encoding schemes are reflected by bounds on $r$ and $R_1$ derived below.


\vspace{3mm}

\noindent{\bf{Encoding:}} 
To define our encoder, we introduce a register $C$ such that $\ket{\rho_{CAS \mid u,v}}$ is a purification of  the state $\rho_{A S \mid u,v}$, which is given in Theorem \ref{achievabilityasy}.
Thus, we have the following purification of the state $\frac{1}{2^{(R_1+r)}} \sum_{(i,j)}\rho_{S \mid U(i)V(m,i, j)}:$ 
\begin{align}
\label{purif}
{\ket{\tau_{CASIJ \mid U(1), \cdots U(2^{r}), V(m, 1,1 ), \cdots, V(m,2^r, 2^{R_1})} }}:= \frac{1}{\sqrt{2^{(R_1+r)}}} \sum_{(i,j)}\ket{\rho_{CAS\mid U(i)V(m,i,j)}} \ket{i}_I\ket{j}_J.
\end{align}\\
It follows from the Uhlmann's theorem (Fact \ref{uhlmann}) that for every $m \in [1:2^{R}],$ there exists  a set of  isometries $\left\{W^{S' \to ACIJ}_{\cC_{U}, \cC_{m}} \right\}$ such that 
\begin{align}
& \Pur \left( \tau_{CASIJ \mid U(1), \cdots U(2^{r}), V(m, 1,1 ), \cdots, V(m,2^r, 2^{R_1})}, W^{S' \to ACIJ}_{\cC_{U}, \cC_{m}}\left(\phi_{{S'S}}\right) W^{\dagger S' \to ACIJ}_{\cC_{U}, \cC_{m}} \right)\nonumber\\
\label{purifuhlmanns}
&= \Pur \left(\frac{1}{2^{n(R_1+r)}} \sum_{(i,j)}\rho_{S \mid U(i)V(m,i, j)}, \rho_S) \right),
\end{align}
where $\rho_{S \mid u,v}:= \Tr_A \rho_{A S\mid u,v} $.
Using these notations, we define our encoder 
depending on the codewords in the codebook $\mathcal{C}_{U}$ and $\{\mathcal{C}_{m}\}_{m\in [1:2^{R}] }$ 
 as follows.
When Alice intends to send the message $m$, 
she applies the isometry $W^{S'\to ACIJ}_{\cC_{U}, \cC_{m}}$ ( obtained in the derivation of \eqref{purifuhlmanns}) on her register $S'$ and transmits the register $A$ across the channel $\cN_{AS \to B}.$

\vspace{3mm}

\noindent{\bf{Pinching:}} Our decoder will be based on the method of pinching. Therefore, before designing our decoder we first discuss this method.
Consider the following classical-quantum states 
 \begin{align}
 \rho_{UVB} :=& \sum_{u,v}p_{UV}(u,v)\ket{u}\bra{u}_U\otimes\ket{v}\bra{v}_V\otimes \rho_{B \mid u,v};\\
 \rho_{V-U-B} := &\sum_{u}p_{U}(u)\ket{u}\bra{u}_U\otimes\rho_{V \mid u}\otimes \rho_{B \mid u},
 \end{align}
where in the above $\rho_{V \mid u}$ and $\rho_{B \mid u}$ are appropriate marginals with respect to the state $\rho_{UVB}.$

In the subsequent discussions the main aim is to come up with completely positive and trace preserving operations such that at the end of these operations the states $\rho_{UVB},$ $\rho_{UV}\otimes \rho_B$ and $\rho_{V-U-B}$ start commuting. Towards this we define the following operations: $\cE_1$ be the pinching operation with respect to the spectral decomposition of the state $\rho_B.$ 
Further, for every $u,$ let $\cE_{2 \mid u}$ be the pinching operation with respect to the spectral decomposition of the operator $\cE_{1}\left(\rho_{B \mid u}\right).$ Then, $\cE_{2}$ is defined as 
$\cE_{2}(\rho):=\sum_{u}  |u\rangle \langle u| \otimes 
\cE_{2 \mid u}( \langle u| \rho|u\rangle)$.
It easy to observe that $\cE_1(\rho_{V-U-B}),$ $\cE_2\left(\rho_{UVB}\right)$ and the state $\rho_{UV} \otimes \rho_B$ commute with each other. 
In what follows further in this section we will use the notation $v_1$ and $v_2$ to represent  
the maximum number of components of the pinching map ${\cal E}_1$ and the maximum number of components of the pinching maps 
$\{{\cal E}_{2 \mid u}\}_{u}$.  
Further, in the discussions below we will define pinching maps $\cE_3,$ $\cE_{4\mid u}$ and $\cE_{5\mid u},$  where $\cE_3$ is the pinching map with respect to the the spectral basis of $\rho_S,$ $\cE_{4\mid u}$ is the pinching map with respect to the spectral basis of the operator $\cE_3 \left( \rho_{S \mid u}\right)$ and $\cE_{5 \mid u}$ is defined with respect to the state $\rho_{E \mid u}.$ Then, $\cE_{4}$ and $\cE_5$ are defined from $\cE_{4\mid u}$ and $\cE_{5\mid u}$
in the same way as $\cE_{2}$.
Further, let $v_1, v_2, v_3, v_4,$ and $v_5$ be defined as follows:  
\beq
\label{pinchconst}
\begin{aligned}
v_1&:= \mbox{ 
distinct components of the pinching map}~ \cE_1;\\
v_2&:= \mbox{maximum number of distinct components of the pinching maps}~ \{{\cal E}_{2 \mid u}\}_{u};\\
v_3&:= \mbox{ 
distinct components of the pinching map}~ \cE_3;\\
v_4&:= \mbox{maximum number of distinct components of the pinching maps} ~\{\cE_{4\mid u}\}_{u};\\
v_5&:= \mbox{maximum number of distinct components of the pinching map}~ \{\cE_{5\mid u}\}_{u}\ .\\
\end{aligned}
 \enq

\vspace{3mm}

\noindent{\bf{Decoding:}}
First, for two Hermitian matrices $A$ and $B$, we define 
the projection $\{A\ge B\}$ as 
$\sum_{j:\lambda_j \ge 0}P_j$, where
the spectral decomposition of $A-B $ is given as $\sum_{j}\lambda_j P_j$.
In this notation, $P_j$ is the projection to the eigenspace corresponding to the eigenvalue $\lambda_j$.
Then, we define the following projectors:
\begin{align}
\Pi_{UVB}(1):=&\{ {\cal E}_2(\rho_{UVB}) \ge 2^{R+R_1+r} \rho_{UV}\otimes \rho_B\},\\
\Pi_{UVB}(2):=&\{ {\cal E}_2(\rho_{UVB}) \ge 2^{R+R_1} {\cal E}_1(\rho_{V-U-B})\}.
\end{align}

Let $\Pi_{UVB}:= \Pi_{UVB}(1) \Pi_{UVB}(2)=  \Pi_{UVB}(2) \Pi_{UVB}(1).$ For every $(m,i,j) \in \left[1:2^R \right] \times \left[1:2^r \right] \times \left[1:2^{R_1} \right]$ define the following operator: 
\beq
\label{gammas}
\gamma(m,i,j):= \Tr_{UV}\left[\Pi_{UVB}\left(\ket{U(i)}\bra{U(i)}_U \otimes \ket{V(m,i,j)}\bra{V(m,i,j)}_V \otimes \mathbb{I}_B\right)\right].
\enq
We now scale these operators to obtain a valid set of POVM operators as follows: 
\beq
\label{betas}
\beta(m,i,j):= \left(\sum_{(m',i',j')}\gamma(m',i',j')\right)^{-\frac{1}{2}}\gamma(m,i,j) \left(\sum_{(m',i',j')}\gamma(m',i',j')\right)^{-\frac{1}{2}}.
\enq
Bob uses the above set of decoding POVM operators to decode the transmitted message.

\vspace{3mm}

\noindent{\bf{Average performance:}} 
Under the above random construction, 
we can evaluate the average performances.
Let$ M$ be the message which was transmitted by Alice using the strategy above and let $\hat{M}$ be the
decoded message by Bob using the decoding POVMs defined in \eqref{betas}. Notice that by the symmetry of the encoding
and decoding strategy, it is enough to bound $\Pr \left\{\hat{M} \neq 1 | M = 1\right\}$. The following lemma discusses the average performance of our protocol. 

\begin{lemma}\label{L1}
The average performances are evaluated with $\alpha \in (0,1)$ as follows.
\begin{align}\label{fes}
\mathbb{E}_{\cC}\Pr \left\{\hat{M} \neq 1 | M = 1\right\} & \le20\left(v_2^\alpha 2^{\alpha\left(R+R_1+r - \underline{\mathrm{D}}_{1-\alpha} (\rho_{UVB}\| \rho_{UV}\otimes \rho_B)
\right) }+ v_2^\alpha 2^{\alpha \left( R+R_1 - \underline{I}_{1-\alpha}[V;B|U ]\right)}  \right) \nonumber\\
&\hspace{5mm} +
{
\frac{2}{\alpha }\left(\left(\frac{ v_3}{2^{r}}\right)^\alpha
2^{\alpha \underline{\mathrm{D}}_{1+\alpha} (\rho_{US} \| \rho_U \otimes \rho_S) }
+\left(\frac{ v_4}{2^{R_1+r}}\right)^\alpha
2^{\alpha \underline{\mathrm{D}}_{1+\alpha} (\rho_{UVS} \| \rho_{UV} \otimes \rho_S) } \right)   
},
\end{align}
and
\begin{align}
\mathbb{E}_{\cC}\left[ \Pur\left( \rho_{ME} ,  \tr_{E}[\rho_{ME}] \otimes  \tr_{M}[\rho_{ME}]\right)^2
\right] 
&\le {
\frac{8}{\alpha }\left(\left(\frac{ v_1}{2^{r}}\right)^\alpha
2^{\alpha \underline{\mathrm{D}}_{1+\alpha} (\rho_{US} \| \rho_U \otimes \rho_S) }
+\left(\frac{ v_2}{2^{R_1+r}}\right)^\alpha
2^{\alpha \underline{\mathrm{D}}_{1+\alpha} (\rho_{UVS} \| \rho_{UV} \otimes \rho_S) } \right)   
} \nonumber\\
&\hspace{5mm}+{
\frac{8}{\alpha }\left(\frac{v_5^\alpha}{2^{\alpha R_1}} 2^{\alpha{\underline{\mathrm{ D}}}_{1+ \alpha} \left(\rho_{UVE} \| \rho_{U-V-E} \right)}\right)    
},\label{avgsec}
\end{align}
where $v_1,v_2,v_3,v_4,$ and $v_5$ are constants defined in \eqref{pinchconst}.

\end{lemma}
This lemma will be proven in Section\ \ref{single-shot-error}. For now we assume this lemma and prove the existence of a code which is robust to both decoding error and secrecy. 
\vspace{3mm}

\noindent{\bf{Existence of good code:}} 
Applying expurgation to this construction, we obtain the following theorem.

\begin{theorem}
\label{sach}
For $\alpha \in (0,1)$ and for every $R,R_1,r >0,$ there exists a code such that 
\begin{align*}
\Pr\left\{ M \neq \hat{M}\right\} & \leq 42\left(v_2^\alpha 2^{\alpha\left(R+R_1+r - \underline{\mathrm{D}}_{1-\alpha} (\rho_{UVB}\| \rho_{UV}\otimes \rho_B)
\right) }+ v_2^\alpha 2^{\alpha \left( R+R_1 - \underline{I}_{1-\alpha}[V;B|U ]\right)}  \right) \nonumber\\
&\hspace{5mm} + 
{
 \frac{5}{\alpha }\left(\left(\frac{ v_3}{2^{r}}\right)^\alpha
2^{\alpha \underline{\mathrm{D}}_{1+\alpha} (\rho_{US} \| \rho_U \otimes \rho_S) }
+\left(\frac{ v_4}{2^{R_1+r}}\right)^\alpha
2^{\alpha \underline{\mathrm{D}}_{1+\alpha} (\rho_{UVS} \| \rho_{UV} \otimes \rho_S) } \right)   
}, 
\end{align*}
and 
\begin{align*}
\Pur\left( \rho_{ME} ,  \tr_{E}[\rho_{ME}] \otimes  \tr_{M}[\rho_{ME}]\right)^2 \le&
20 \bigg({
\frac{1}{\alpha }\left(\left(\frac{ v_3}{2^{r}}\right)^\alpha
2^{\alpha \underline{\mathrm{D}}_{1+\alpha} (\rho_{US} \| \rho_U \otimes \rho_S) }
+\left(\frac{ v_4}{2^{R_1+r}}\right)^\alpha
2^{\alpha \underline{\mathrm{D}}_{1+\alpha} (\rho_{UVS} \| \rho_{UV} \otimes \rho_S) } \right)   
}\\
&\hspace{2mm}+ {
\frac{1}{\alpha }\left(\frac{v_5^\alpha}{2^{\alpha R_1}} 2^{\alpha{\underline{\mathrm{ D}}}_{1+ \alpha} \left(\rho_{UVE} \| \rho_{V-U-E} \right)}\right)    
},\bigg.
\end{align*}
where $v_,v_2,v_3,v_4,$ and $v_5$ are constants defined in \eqref{pinchconst} and the information theoretic quantities above are calculated with respect to the state $\rho_{UVAS}=\sum_{(u,v)}p_{UV}(u,v) \ket{u}\bra{u} _U\otimes \ket{v}\bra{v}_V \otimes \rho_{AS \mid u,v}$ such that $\tr_{UVA}{\rho_{UVAS}}= \tr_{S}\phi_{S'S}.$ 
\end{theorem}

\begin{proof}
We now show the existence of a code which simultaneously satisfies both the reliability and the secrecy criterion as discussed in the Definition \eqref{edes}. Towards this let $\eps\left(\cC\right)$ and  $\delta(\cC)$ represent the decoding error and secrecy parameter of a random codebook $\cC$. Define the following events:
\begin{align}
\label{e1}
\cB_1 &:= \left\{ \eps(\cC) \leq {(1+\beta)}\mathbb{E}_{\cC}[\eps(\cC)] \right\};\\
\label{e2}
\cB_2 &:= \left\{ \delta(\cC) \leq {(1+\beta)}\mathbb{E}_{\cC}[\delta(\cC)]\right\},
\end{align}
where $\beta > 1$ is an arbitrary constant. 
From Markov's inequality and union bound it now easily follows that
\beq
\label{markov}
\Pr\left\{\cB_1, \cB_2\right\} \geq \frac{\beta - 1}{ \beta + 1}.
\enq
Thus, from \eqref{fes}, \eqref{avgsec}, \eqref{e1},  \eqref{e2} and setting  $\beta= 1.1$ in \eqref{markov} we now conclude that there exists a codebook such that:
\begin{align*}
\Pr\left\{ M \neq \hat{M}\right\} 
& \leq 42 \left(v_2^\alpha 2^{\alpha\left(R+R_1+r - \underline{\mathrm{D}}_{1-\alpha} (\rho_{UVB}\| \rho_{UV}\otimes \rho_B)
\right) }+ v_2^\alpha 2^{\alpha \left( R+R_1 - \underline{I}_{1-\alpha}[V;B|U ]\right)}  \right) \nonumber\\
&\hspace{5mm} +
 {
 \frac{5}{\alpha }\left(\left(\frac{ v_3}{2^{r}}\right)^\alpha
2^{\alpha \underline{\mathrm{D}}_{1+\alpha} (\rho_{US} \| \rho_U \otimes \rho_S) }
+\left(\frac{ v_4}{2^{R_1+r}}\right)^\alpha
2^{\alpha \underline{\mathrm{D}}_{1+\alpha} (\rho_{UVS} \| \rho_{UV} \otimes \rho_S) } \right)   
}, 
\end{align*}
and 
\begin{align*}
\Pur\left( \rho_{ME} ,  \tr_{E}[\rho_{ME}] \otimes  \tr_{M}[\rho_{ME}]\right)^2 \le& 
20\bigg({
\frac{1}{\alpha }\left(\left(\frac{ v_3}{2^{r}}\right)^\alpha
2^{\alpha \underline{\mathrm{D}}_{1+\alpha} (\rho_{US} \| \rho_U \otimes \rho_S) }
+\left(\frac{ v_4}{2^{R_1+r}}\right)^\alpha
2^{\alpha \underline{\mathrm{D}}_{1+\alpha} (\rho_{UVS} \| \rho_{UV} \otimes \rho_S) } \right)   
}\\
&\hspace{5mm}+ {
\frac{1}{\alpha }\left(\frac{v_5^\alpha}{2^{\alpha R_1}} 2^{\alpha{\underline{\mathrm{ D}}}_{1+ \alpha} \left(\rho_{UVE} \| \rho_{U-V-E} \right)}\right)   
}\bigg).
\end{align*}
This completes the proof. 
\end{proof}

\section{Asymptotic analysis}
\subsection{Preparation}
To analyze the asymptotic case,
first we bound the number of distinct components of the pinching maps $\cE_1$ and $\cE_2$ in the asymptotic and i.i.d. case.
That is, we consider the case when there are $n$ independent copies of the states $\rho_{UVB}$ and $\rho_{V-U-B}.$ 
Let $d_U$ and $d_B$ be the dimensions of $\cH_U$ and $\cH_B$. The lemma below gives an upper bound on the number of distinct components of the maps $\cE_1$ and $\cE_2 $ discussed above.
\begin{lemma}\label{componentbound}
Let $\cE_1$ and $\cE_2$ be the pinching maps as defined above. Further, let $v_1, v_2$ represent the number of distinct components of the map $\cE_1$ and $\cE_2$ respectively. Then,
\begin{align*}
v_1 \le (n+1)^{d_B-1},~ v_2 \le (n+1)^{d_U(d_B+2)(d_B-1)/2}.
\end{align*}
\end{lemma}
\begin{proof}
$\rho_B^{\otimes n}$ has $(n+1)^{d_B-1}$ eigenvalues at most.
Hence, $v_1 \le (n+1)^{d_B-1}$.

We now prove the upper bound on $v_2.$ Towards this let $\{|u_1\rangle, \ldots, |u_{d_U}\rangle\}$ represent the basis of $\mathcal{H}_U.$ We now focus on the the number of components of the pinching map ${\cal E}_{2|\vec{u}},$ where $\vec{u}:= (\underbrace{u_1,\ldots, u_1}_{n_1}, \ldots \underbrace{u_{d_U},\ldots,u_{d_U} }_{n_{d_U}})$.
The state ${\rho}_{B^n|\vec{u}}$
is written as
$ \rho_{B \mid u_1}^{ \otimes n_1}\otimes \cdots \otimes \rho_{B \mid u_{d_B}}^{ \otimes n_{d_B}}$.
Then, the space $\cH_B^{\otimes n_j}$ is decomposed to 
\begin{align}
\cH_B^{\otimes n_j}=\bigoplus_{\lambda \in Y_{d}^{n_j}} \cU_\lambda(\SU) \otimes \cU_\lambda (S_{n_j}),
\end{align}
where $Y_d^{n_j}$ is the set of indexes of size $n_j$ and depth not greater than $ d_B$. 
We have
$|Y_{d_B}^{n_j}| \le n_j^{d_B-1} $
and 
Weyl's dimension formula shows that
$\dim \cU_\lambda(\SU) \le (n+1)^{d_B(d_B-1)/2}$.

We denote the pinching whose components are 
$\{ \cU_\lambda(\SU) \otimes \cU_\lambda (S_{n_j})\}
_{\lambda \in Y_{d}^{n_j}}$
by $\cE_{n_j}$.
The states ${\rho}_{B|\vec{u}}$
and $\rho_B^{\otimes n}$
are invariant with respect to
$\cE_{n_1} \otimes \cdots \otimes \cE_{n_{d_U}}$.
Therefore, 
$\cE_1(\rho_{V-U-B})=
\cE_1\circ \cE_{n_1} \otimes \cdots \otimes \cE_{n_{d_B}}(\rho_{V-U-B})$.
 
Now, we consider each component of 
$\cE_{n_1} \otimes \cdots \otimes \cE_{n_{d_U}}$.
That is, we consider the subspace
$\cU_{\lambda_1}(\SU) \otimes \cU_{\lambda_1} (S_{n_1})
\otimes \cdots \otimes
\cU_{\lambda_{d_U}}(\SU) \otimes \cU_{\lambda_{d_U}} 
(S_{n_{d_U}})
=(\cU_{\lambda_1}(\SU) 
\otimes \cdots \otimes \cU_{\lambda_{d_B}}(\SU) )
\otimes 
(\cU_{\lambda_1} (S_{n_1}) \otimes \cdots 
\otimes \cU_{\lambda_{d_U}} (S_{n_{d_U}}))
$.
Both states are the identity on 
$\cU_{\lambda_1} (S_{n_1}) \otimes \cdots 
\otimes \cU_{\lambda_{d_U}} (S_{n_{d_U}})$.
Thus, 
on this subspace,
the number of eigenvalues of
$\cE_1(\rho_{V-U-B})$ is 
the dimension of 
$\cU_{\lambda_1}(\SU) \otimes \cdots \otimes \cU_{\lambda_{d_U}}(\SU) $
at most. 
The dimension is $(n+1)^{d_Ud_B(d_B-1)/2}$ at most.
Further, the number of components of 
$\cE_{n_1} \otimes \cdots \otimes \cE_{n_{d_B}}$
is $(n+1)^{d_B(d_U-1)}$ at most.
Therefore, the number of eigenvalues of
$\cE_1(\rho_{V-U-B})$ is 
$(n+1)^{d_Ud_B(d_B-1)/2}
(n+1)^{d_U(d_B-1)}=
(n+1)^{d_U(d_B+2)(d_B-1)/2}$ at most.
\end{proof}

Further, we have the following additivity property.
\begin{lemma}
\label{additivity}
\begin{align}
\underline{I}_{1-\alpha}[V; B \mid U]_{\rho_{UVB}^{\otimes n}|\rho_{UV}^{\otimes n}}  
=
n \underline{I}_{1-\alpha}[V;B \mid U]_{\rho_{UVB}|\rho_{UV}}.
\label{L33}
\end{align}
\end{lemma}
\begin{proof}
Let $\sigma_{UVB}$ be the state which attains the minimum in the definition of $\min_{\sigma_{B}}\underline{\mathrm{{D}}}_{1+\alpha}(\rho_{UVB}\|\rho_{UV}$ (see \eqref{Min11})
It then follows from the uniqueness condition that (see \eqref{condis}), $\sigma_{UVB}^{\otimes n}$ satisfies
the condition \eqref{condis} for the $n$-copy case. The statement of the lemma now follows from the uniqueness condition.
Due to the uniqueness condition, we obtain \eqref{L33}.
\end{proof}

\subsection{Proof of Theorem \ref{achievabilityasy}}
Now, we proceed to our proof for Theorem \ref{achievabilityasy}.
From Theorem \ref{sach} it is easy to see that if the channel $\cN_{AS \to BE}$ is used $n$ times independently, then there exists a code such that 
\begin{align}
&\Pr\left\{ M \neq \hat{M}\right\}  \nonumber\\
\leq & 42 \left(v_2^\alpha 2^{\alpha\left(n(R+R_1+r) - \underline{\mathrm{D}}_{1-\alpha} (\rho^{\otimes n}_{UVB}\| \rho^{\otimes n}_{UV}\otimes \rho^{\otimes n}_B)
\right) }+ v_2^\alpha 2^{\alpha \left(n (R+R_1) - \underline{I}_{1-\alpha}[V^n;B^n|U^n ]_{\rho^{\otimes n}_{UVB} \mid \rho^{\otimes n}_{UV}}\right)}  \right) \nonumber\\
&\hspace{5mm} + 
 \frac{5}{\alpha }\left(\left(\frac{ v_3}{2^{nr}}\right)^\alpha
2^{\alpha \underline{\mathrm{D}}_{1+\alpha} (\rho^{\otimes n}_{US} \| \rho^{\otimes n}_U \otimes \rho^{\otimes n}_S) }
+\left(\frac{ v_4}{2^{n(R_1+r)}}\right)^\alpha
2^{\alpha \underline{\mathrm{D}}_{1+\alpha} (\rho^{\otimes n}_{UVS} \| \rho^{\otimes n}_{UV} \otimes \rho^{\otimes n}_S) }    
\right),
\label{asym11}
\end{align}
and 
\begin{align}
\Pur\left( \rho^{\otimes n}_{ME} ,  \tr_{E}[\rho^{\otimes n}_{ME}] \otimes  \tr_{M}[\rho^{\otimes n}_{ME}]\right)^2 
\le& 
20 \bigg({
 \frac{1}{\alpha}\left(\left(\frac{ v_3}{2^{nr}}\right)^\alpha
2^{\alpha \underline{\mathrm{D}}_{1+\alpha} (\rho^{\otimes n}_{US} \| \rho^{\otimes n}_U \otimes \rho^{\otimes n}_S) }
+\left(\frac{ v_4}{2^{n(R_1+r)}}\right)^\alpha
2^{\alpha \underline{\mathrm{D}}_{1+\alpha} (\rho^{\otimes n}_{UVS} \| \rho^{\otimes n}_{UV} \otimes \rho^{\otimes n}_S) } \right)   
} \nonumber\\
\label{asym2}
&\hspace{5mm}+ {
\frac{1}{\alpha }\left(\frac{v_5^\alpha}{2^{\alpha nR_1}} 2^{\alpha{\underline{\mathrm{ D}}}_{1+ \alpha} \left(\rho^{\otimes n}_{UVE} \| \rho^{\otimes n}_{V-U-E} \right)}\right)    
} \bigg).
\end{align}
Hence, using Lemma \ref{componentbound} and Lemma \ref{additivity}, we have
\begin{align}
&\lim_{n \to \infty }\frac{-1}{n}\log \Pr\left\{ M \neq \hat{M}\right\}  \nonumber\\
\ge &
\min \Big( \alpha \underline{\mathrm{D}}_{1-\alpha} (\rho_{UVB}\| \rho_{UV}\otimes \rho_B)
-(R+R_1+r), \nonumber\\
&\alpha \left(\underline{I}_{1-\alpha}[V;B|U]_{\rho_{UVB} \mid \rho_{UV}}- (R+R_1) \right) ,\nonumber\\
&\alpha \big( r- \underline{\mathrm{D}}_{1+\alpha} (\rho_{US} \| \rho_U \otimes \rho_S) \big), \nonumber\\
&\alpha \big((R_1+r)- \underline{\mathrm{D}}_{1+\alpha} (\rho_{UVS} \| 
\rho_{UV} \otimes \rho_S)\big)\Big),\label{asym11A}
\end{align}
and
\begin{align}
&\lim_{n \to \infty }\frac{-1}{n}\log \Pur\left( \rho^{\otimes n}_{ME} ,  \tr_{E}[\rho^{\otimes n}_{ME}] \otimes  \tr_{M}[\rho^{\otimes n}_{ME}]\right)  \nonumber\\
\ge &
\min \Big( 
\alpha \big( r- \underline{\mathrm{D}}_{1+\alpha} (\rho_{US} \| \rho_U \otimes \rho_S) \big), \nonumber\\
&\alpha \big((R_1+r)- \underline{\mathrm{D}}_{1+\alpha} (\rho_{UVS} \| 
\rho_{UV} \otimes \rho_S)\big), \nonumber\\
&\alpha \big(R_1 -
\underline{\mathrm{ D}}_{1+ \alpha} 
\left(\rho_{UVE} \| \rho_{V-U-E} \right) \big)
\Big). \label{asym12A}
\end{align}
Thus, it now follows from \eqref{asym11A}, \eqref{asym12A}
that as $n \to \infty$ and $\alpha \to 0,$ then there exists a code such that 
\begin{align*}
\lim_{n\to \infty}\Pr\left\{ M \neq \hat{M}\right\} &\to 0,\\
\lim_{n\to \infty}\Pur\left( \rho^{\otimes n}_{ME} ,  \tr_{E}[\rho^{\otimes n}_{ME}] \otimes  \tr_{M}[\rho^{\otimes n}_{ME}]\right) &\to 0; 
\end{align*}
if,
\begin{align}
\label{rb1}
R+R_1+r & < I[UV;B];\\
\label{rb2}
R+R_1 & < I[V;B \mid U];\\
\label{rb3}
R_1+ r & > I[UV;S];\\
\label{rb4}
r & >I[U;S];\\
\label{rb5}
R_1 &> I [V;E \mid U].
\end{align}  
Now, for arbitrary small real numbers $\epsilon_1,\epsilon_2,\epsilon_3>0$,
we set 
$R_1:= I[V;E \mid U]+\epsilon_1$
and $r:=\max( I[U;S],I[UV;S]- I[V;E \mid U])+\epsilon_2$,
which implies that $R_1+r=\max( I[U;S]+I[V;E \mid U],I[UV;S] )
+\epsilon_1+\epsilon_2$.
Then, we set $R:= \min(I[UV;B]- (R_1+r), I[V;B \mid U]-R)-\epsilon_3$.
With this choice, the aforementioned conditions are satisfied.
In this case, the rate $R$ can be written as
\begin{align}
& R=\min(I[UV;B]- (R_1+r), I[V;B \mid U]-R)-\epsilon_3 \nonumber \\
=&\min \Big(I[UV;B]- 
\max( I[U;S]+I[V;E \mid U],I[UV;S] ) -\epsilon_1-\epsilon_2,\nonumber \\
&I[V;B \mid U]-I[V;E \mid U]-\epsilon_1\Big)-\epsilon_3\nonumber\\
=&\min \Big(
I[UV;B]- I[V;E \mid U]- I[U;S]-\epsilon_1-\epsilon_2, \nonumber\\
&I[UV;B]- I[UV;S]-\epsilon_1-\epsilon_2 ,\nonumber\\
&I[V;B \mid U]-I[V;E \mid U]-\epsilon_1\Big)-\epsilon_3.\label{MMH}
\end{align}
Since $\epsilon_1,\epsilon_2,\epsilon_3$ are arbitrary small real numbers,
the rate given in \eqref{NHFG} is achievable.
This completes the proof of Theorem \ref{achievabilityasy}.

\section{Proof of Lemma \ref{L1}} 
\label{single-shot-error}
\subsection{Error Analysis}
First, we show \eqref{fes} of Lemma \ref{L1}.
Our proof employs Lemmas \ref{decerr} and \ref{bcove}, which are given in latter sections.
Towards this let $\Theta_{B}(1)$ be the state received by the Bob when Alice 
transmits the message $m=1.$ 
Hence, it is given as $ \tr_{CEIJ} \cN_{AS\to BE } 
\left(W^{S' \to ACIJ}_{\cC_{U}, \cC_{1}}\left(\phi_{{S'S}}\right) 
W^{\dagger S' \to ACIJ}_{\cC_{U}, \cC_{1}} \right)$.
Further, let $\hat{\Theta}_{B}(1)$ be defined as follows:
\beq
\hat{\Theta}_{B}(1):= \frac{1}{2^{(R_1+r)}} \sum_{(i,j) \in [1:2^{r}] \times [1:2^{R_1}]} \tr_{CE} \cN_{AS\to BE } \left(\rho_{CAS \mid U(i)V(m,i,j)} \right). \label{Theta}
\enq
We now bound $\Pr \left\{\hat{M} \neq 1 | M = 1\right\}$ average over the random choice of the codebook. In what follows we will use the notation $\cC$ to denote the random choice of sequences mentioned in the codebook above. The error is now bounded as follows: 
\begin{align}
&\mathbb{E}_{\cC}\Pr \left\{\hat{M} \neq 1 | M = 1\right\}  \nonumber\\
 = &\mathbb{E}_{\cC}
\left[\tr \left[  \left(\sum_{(m' \neq1 ), k ,l}\beta(m',k,l) \right)\Theta_{B}(1)\right]\right] \nonumber\\
\overset{a} \leq  &2 \mathbb{E}_{\cC} \left[\tr \left[  \left(\sum_{(m' \neq1 ), k ,l}\beta(m',k,l) \right)\hat{\Theta}_{B}(1)\right]\right] 
 \nonumber\\
 &+  2 \mathbb{E}_{\cC}
\left|
\sqrt{\left[\tr \left[  \left(\sum_{(m' \neq1 ), k ,l}\beta(m',k,l) \right)\Theta_{B}(1)\right]\right]}
-
\sqrt{\left[\tr \left[  \left(\sum_{(m' \neq1 ), k ,l}\beta(m',k,l) \right)\hat{\Theta}_{B}(1)\right]\right]}
\right|^2
\nonumber\\
\overset{b} \leq &2 \mathbb{E}_{\cC} \left[\tr \left[  \left(\sum_{(m' \neq1 ), k ,l}\beta(m',k,l) \right)\hat{\Theta}_{B}(1)\right]\right] +  2 \mathbb{E}_{\cC}\left[\Pur({\Theta}_{B}(1), \hat{\Theta}_{B}(1))^2\right],
\label{splitstates}
\end{align}
where 
$a$ follows from the generic inequality $(x+y)^2 \leq 2(x^2 + y^2)$; 
$b$ follows from the Fact \ref{err} and the facts that $\sum_{(m' \neq1 ), k ,l}\beta(m',k,l) \preceq \mathbb{I}$.

Using 
$\tau_{CAS \mid \cC} :=\frac{1}{2^{(R_1+r)}} \sum_{(i,j) \in [1:2^{r}] \times [1:2^{R_1}]} \rho_{CAS \mid U(i)V(1,i,j)}
$
and
$\tau_{S \mid \cC} :=\frac{1}{2^{(R_1+r)}} \sum_{(i,j) \in [1:2^{r}] \times [1:2^{R_1}]} 
\Tr_{CA} \rho_{CAS \mid U(i)V(1,i,j)}
$,
we will first bound the second term in \eqref{splitstates} as follows: 
\begin{align}
&2 \mathbb{E}_{\cC}\left[\Pur(
 \hat{\Theta}_{B}(1),
{\Theta}_{B}(1)
)^2\right]\nonumber \\
&= 2 \mathbb{E}_{\cC}\bigg[\Pur\bigg(\frac{1}{2^{(R_1+r)}} \sum_{(i,j) \in [1:2^{r}] \times [1:2^{R_1}]} 
\tr_{CE} \cN_{AS\to BE } \left(\rho_{CAS \mid U(i)V(m,i,j)} \right), \nonumber\\
&\hspace{15mm}\tr_{CEIJ} \cN_{AS\to BE } \left(W^{S' \to ACIJ}_{\cC_{U}, \cC_{1}}\left(\phi_{{S'S}}\right) W^{\dagger S' \to ACIJ}_{\cC_{U}, \cC_{1}} \right)\bigg)^2\bigg] \nonumber\\
&\overset{a} \leq 2 \mathbb{E}_{\cC}\bigg[\Pur\bigg(
\tau_{CAS \mid \cC}
, W^{S' \to ACIJ}_{\cC_{U}, \cC_{1}}\left(\phi_{{S'S}}\right) 
W^{\dagger S' \to ACIJ}_{\cC_{U}, \cC_{1}}
\bigg)^2\bigg] \nonumber\\
&\overset{b}= 
2 \mathbb{E}_{\cC}\bigg[\Pur \left(\tau_{S \mid \cC}, \rho_S \right)^2\bigg] \nonumber\\
&\overset{c}\le 
 2 \mathbb{E}_{\cC}\bigg[
1- 2^{\alpha \underline{\mathrm{D}}_{1+\alpha} (\tau_{S \mid \cC} \|  \rho_S) }
\bigg]\nonumber\\
&\overset{d}\le 
 2{\ln 2 \cdot \mathbb{E}_{\cC}\left[\underline{\mathrm{D}}_{1+\alpha}\left(\tau_{S \mid \cC} \| \rho_S\right) \right]}
\nonumber\\
&\overset{e}\le 
{
 \frac{2}{\alpha }\left(\left(\frac{ v_3}{2^{r}}\right)^\alpha
2^{\alpha \underline{\mathrm{D}}_{1+\alpha} (\rho_{US} \| \rho_U \otimes \rho_S) }
+\left(\frac{ v_4}{2^{R_1+r}}\right)^\alpha
2^{\alpha \underline{\mathrm{D}}_{1+\alpha} (\rho_{UVS} \| \rho_{UV} \otimes \rho_S) } \right)   
} 
\label{int11}
\end{align}
where $a$ follows from Fact \ref{fact:monotonequantumoperation} 
with respect to the map $\tr_{CE} \cN_{AS\to BE }$;
$b$ follows from \eqref{purifuhlmanns};
$c$ follows from Fact \ref{pinsker};
$d$ follows from the following relation \eqref{EH1}; 
and 
$e$ follows from Lemma \ref{bcove}.
\begin{align}
{1-2^{-\frac{x}{\ln 2}}}=
{1-e^{-x}}\le x.\label{EH1}
\end{align}

We now bound the first term on the R.H.S of \eqref{splitstates} 
by using several steps as follows.
\begin{align}
& \sum_{(m' \neq1 ), k ,l}\mathbb{E}_{\cC} \left[\tr \left[ \beta(m',k,l) \hat{\Theta}_{B}(1)\right]\right] \nonumber\\
  & = \frac{1}{2^{(R_1+r)}} \sum_{i,j} \sum_{(m'\neq 1),k,l} \mathbb{E}_{\cC}\left[\tr \left[\beta \left(m',k ,l \right) \rho_{B \mid U(i)V(1,i,j)}\right]\right] \nonumber\\
&\overset{a} = \sum_{(m'\neq 1),k,l} \mathbb{E}_{\cC}\left[\tr \left[\beta \left(m',k ,l \right) \rho_{B \mid U(1)V(1,1,1)}\right]\right] \nonumber\\
& \leq \mathbb{E}_{\cC}\left[ \tr\left[\left( \mathbb{I} -\beta\left(1,1,1\right) \right)\rho_{B \mid U(1)V(1,1,1)} \right]\right] \nonumber \\
\label{intermediatetermss}
& \overset{b} \leq 2\mathbb{E}_{\cC}\left[ \tr\left[\left( \mathbb{I} -\gamma \left(1,1,1\right) \right)\rho_{B \mid U(1)V(1,1,1)} \right]\right] + 4 \sum_{(m',k,l) \neq (1 ,1,1)}\mathbb{E}_{\cC}\left[ \tr \left[ \gamma \left(m',k,l\right)\rho_{B \mid U(1)V(1,1,1)}  \right]\right],
\end{align}
where $a$ follows from the symmetry of the code construction and $b$ follows from the Hayashi-Nagaoka operator inequality ( Fact \ref{haynag}). We now bound each of the terms on the right hand side of \eqref{intermediatetermss}. \par
Consider $2\mathbb{E}_{\cC}\left[ \tr\left[\left( \mathbb{I} -\gamma \left(1,i,j\right) \right)\rho_{B \mid U(i)V(1,i,j)} \right]\right]:$
\begin{align}
&2 \cdot \mathbb{E}_{\cC}\left[ \tr\left[\left( \mathbb{I} -\gamma \left(1,i,j\right) \right)\rho_{B \mid U(1)V(1,1,1)} \right]\right] \nonumber\\
&\overset{a}= 2\cdot \mathbb{E}_{\cC}\left[ \tr\left[\left( \mathbb{I} -\Tr_{UV}\left[\Pi_{UVB}\left(\ket{U(1)}\bra{U(i)}_U \otimes \ket{V(1,1,1)}\bra{V(1,1,1)}_V \otimes \mathbb{I}\right)\right] \right)\rho_{B \mid U(1)V(1,1,1)} \right]\right]\nonumber\\
&\overset{b}=2\cdot \tr\left[ \left(\mathbb{I}- \Pi_{UVB}\right)\rho_{UVB}\right] \nonumber\\
&\overset{c} \leq 2\cdot \tr\left[ \left(\mathbb{I}- \Pi_{UVB}(1)\right)\rho_{UVB}\right] + 2\tr\left[ \left(\mathbb{I}- \Pi_{UVB}(1)\right)\rho_{UVB}\right] \nonumber\\
&\overset{d} \leq 2 \cdot v_2^\alpha 2^{\alpha\left(R+R_1+r\right)} 
2^{-\alpha \underline{\mathrm{D}}_{1-\alpha} (\rho_{UVB}\| \rho_{UV}\otimes \rho_B) } + 2 \cdot v_2^\alpha 2^{\alpha \left( R+R_1\right)} 
\label{term11s}
2^{-\alpha \underline{I}_{1-\alpha}[V;B|U ]_{\rho_{UVB}|\rho_{UV}}},
\end{align} 
where $a$ follows from the definition of $\gamma(1,i,j)$ mentioned in \eqref{gammas}; $b$ follows from the linearity of the trace operation; $c$ follows from the definition of $\Pi_{UVB}$ and the fact $\mathbb{I} - \Pi_{UVB} \preceq \mathbb{I} - \Pi_{UVB}(1) + \mathbb{I} - \Pi_{UVB}(2)$ and $d$ follows from \eqref{ineq3} and \eqref{ineq4} proven in Lemma \ref{decerr}. 

We now bound the second term on the right hand side of \eqref{intermediatetermss} as follows: 
\begin{align}
&4 \cdot \sum_{(m',k,l) \neq (1,1,1)}\mathbb{E}_{\cC}\left[ \tr \left[ \gamma \left(m',k,l\right)\rho_{B \mid U(1)V(1,1,1)}  \right]\right] \nonumber \\
&=4 \cdot   \sum_{k \neq 1 } \mathbb{E}_{\cC}\left[ \tr \left[ \gamma \left(1,k ,1\right)\rho_{B \mid U(1)V(1,,1,1)}  \right]\right] + 4 \cdot \sum_{(m',l) \neq (1,1) } \mathbb{E}_{\cC}\left[ \tr \left[ \gamma \left(m',1,l\right)\rho_{B \mid U(1)V(1,1,1)}  \right]\right]   \nonumber\\
\label{errors2s}
&\hspace{4mm} +4 \cdot  \sum_{m' \neq 1, k \neq 1, l \neq 1}\mathbb{E}_{\cC}\left[ \tr \left[ \gamma \left(m',k,l\right)\rho_{B \mid U(1)V(1,1,1)}  \right]\right].
\end{align}
We now bound each of the terms on the right hand side of \eqref{errors2s}. Consider $\sum_{k \neq i } \mathbb{E}_{\cC}\left[ \tr \left[ \gamma \left(1,k ,j\right)\rho_{B \mid U(i)V(1,i,j)}  \right]\right] :$
\begin{align}
&4 \cdot \sum_{k \neq 1 } \mathbb{E}_{\cC}\left[ \tr \left[ \gamma \left(1,k ,1\right)\rho_{B \mid U(1)V(1,1,1)}  \right]\right] \nonumber \\ 
&\overset{a}=4 \cdot \sum_{k \neq 1 } \mathbb{E}_{\cC}\left[ \tr \left[ \Tr_{UV}\left[\Pi_{UVB}\left(\ket{U(k)}\bra{U(k)}_U \otimes \ket{V(1,k,1)}\bra{V(1,k,1)}_V \otimes \mathbb{I}\right)\right]\rho_{B \mid U(1)V(1,1,1)}  \right]\right] \nonumber\\
&\overset{b}=4 \cdot  2^r \Tr \left[\Pi_{UVB} \rho_{UV} \otimes \rho_{B}\right] \nonumber\\
&\overset{c} \leq 4 \cdot 2^r \Tr \left[\Pi_{UVB}(1) \rho_{UV} \otimes \rho_{B}\right] \nonumber\\
&\overset{d}\leq4 \cdot 2^r v_2^\alpha 2^{-(1-\alpha)(R+r+R_1)} 
2^{-\alpha \underline{\mathrm{D}}_{1-\alpha} (\rho_{UVB}\| \rho_{UV}\otimes \rho_B) } \nonumber\\
\label{errexp1}
&\leq 4 \cdot v_2^\alpha 2^{\alpha\left((R+r+R_1) -  \underline{\mathrm{D}}_{1-\alpha} (\rho_{UVB}\| \rho_{UV}\otimes \rho_B)  \right)} ,
\end{align}
where $a$ follows from the definition of $\gamma(1,k,1)$ mentioned in \eqref{gammas}; $b$ follows from the independence of the random variables involved, linearity of trace operation, from the definition of the states $\rho_{UV}$ and $\rho_B$ and by the symmetry of the code construction $c$ follows because $\Pi_{UVB} \preceq \Pi_{UVB}(1)$ and $d$ follows from \eqref{ineq1} proven in Lemma \ref{decerr}.

We now bound the second term on the right hand side of \eqref{errors2s} as follows: 
\begin{align}
4 \cdot &\sum_{(m',l) \neq (1,1) } \mathbb{E}_{\cC}\left[ \tr \left[ \gamma \left(m',1,l\right)\rho_{B \mid U(1)V(1,1,1)}  \right]\right]   \nonumber \\
& \overset{a} = 4 \cdot \sum_{(m',l) \neq (1,1) } \mathbb{E}_{\cC}\left[ \tr \left[ \Tr_{UV}\left[\Pi_{UVB}\left(\ket{U(1)}\bra{U(1)}_U \otimes \ket{V(m',1,l)}\bra{V(m',1,l)}_V \otimes \mathbb{I}\right)\right]\rho_{B \mid U(1)V(1,1,1)}  \right]\right] \nonumber\\
&\overset{b}=4 \cdot  2^{R+R_1} \Tr \left[\Pi_{UVB} \rho_{V-U-B}\right] \nonumber\\
&\overset{c}\leq 4 \cdot 2^{R+R_1} \Tr \left[\Pi_{UVB}(2) \rho_{V-U-B}\right] \nonumber\\
& \overset{d}\leq4 \cdot  2^{R+R_1} v_2^\alpha2^{-(1-\alpha)(R+R_1)} 2^{-\alpha 
\underline{I}_{1-\alpha}[V;B|U ]} \nonumber\\
\label{errexp2}
& = 4 \cdot v_2^\alpha 2^{\alpha \left( \left(R+R_1\right) - \underline{I}_{1-\alpha}[V;B|U ] \right)},
\end{align}
where $a$ follows from the definition of $\gamma(m',i,l)$ mentioned in \eqref{gammas}; $b$ follows from the independence of the random variables involved, linearity of trace operation and from the definition of the state $\rho_{V-U-B}$ and from the symmetry of the code construction; $c$ follows because $\Pi_{UVB} \preceq \Pi_{UVB}(2)$ $d$ follows from \eqref{ineq2} proven in Lemma \ref{decerr}.

The third term on the right hand side of \eqref{errors2s} is bounded as follows: 
\beq
\label{errexp3}
4 \cdot \sum_{m' \neq 1, k \neq 1, l \neq 1}\mathbb{E}_{\cC}\left[ \tr \left[ \gamma \left(m',k,l\right)\rho_{B \mid U(1)V(1,1,1)}  \right]\right] \leq  4 \cdot  v_2^\alpha 2^{\alpha\left((R+r+R_1) -  \underline{\mathrm{D}}_{1-\alpha} (\rho_{UVB}\| \rho_{UV}\otimes \rho_B)  \right)}.
\enq
The proof for \eqref{errexp3} follows using exactly similar steps and techniques as that used in the proof of \eqref{errexp1}. 

Combining with the above discussion, we now obtain an upper bound for the first term on the R.H.S of \eqref{splitstates}; 
\begin{align}
& \overset{a} \leq 2 \mathbb{E}_{\cC} \left[\tr \left[ \left( \sum_{(m' \neq1 ), k ,l}\beta(m',k,l) \right)\hat{\Theta}_{B}(1)\right]\right] ,\nonumber\\
&\overset{b} \leq 4\mathbb{E}_{\cC}\left[ \tr\left[\left( \mathbb{I} -\gamma \left(1,1,1\right) \right)\rho_{B \mid U(1)V(1,1,1)} \right]\right] + 8 \sum_{(m',k,l) \neq (1 ,1 , 1)}\mathbb{E}_{\cC}\left[ \tr \left[ \gamma \left(m',k,l\right)\rho_{B \mid U(1)V(1,1,1)}  \right]\right] \nonumber \\
& \overset{c} \leq 4\bigg( v_2^\alpha 2^{\alpha\left(R+R_1+r\right)} 
2^{-\alpha \underline{\mathrm{D}}_{1-\alpha} (\rho_{UVB}\| \rho_{UV}\otimes \rho_B) } +  v_2^\alpha 2^{\alpha \left( R+R_1\right)} 
2^{-\alpha \underline{I}_{1-\alpha}[V;B|U ]_{\rho_{UVB}|\rho_{UV}}}\bigg) \nonumber\\
&\hspace{5mm} + 8 \sum_{(m',k,l) \neq (1 ,1 , 1)}\mathbb{E}_{\cC}\left[ \tr \left[ \gamma \left(m',k,l\right)\rho_{B \mid U(1)V(1,1,1)}  \right]\right] 
\nonumber \\
&\overset{d} \leq 4\bigg( v_2^\alpha 2^{\alpha\left(R+R_1+r\right)} 
2^{-\alpha \underline{\mathrm{D}}_{1-\alpha} (\rho_{UVB}\| \rho_{UV}\otimes \rho_B) } +  v_2^\alpha 2^{\alpha \left( R+R_1\right)} 
2^{-\alpha \underline{I}_{1-\alpha}[V;B|U ]_{\rho_{UVB}|\rho_{UV}}}\bigg) \nonumber\\
&\hspace{5mm}+8 \left( 2 \cdot v_2^\alpha 2^{\alpha\left(R+R_1+r\right)} 
2^{-\alpha \underline{\mathrm{D}}_{1-\alpha} (\rho_{UVB}\| \rho_{UV}\otimes \rho_B) } +  v_2^\alpha 2^{\alpha \left( R+R_1\right)} 
2^{-\alpha \underline{I}_{1-\alpha}[V;B|U ]_{\rho_{UVB}|\rho_{UV}}}\right)
\nonumber \\
&\leq 20\left(v_2^\alpha 2^{\alpha\left(R+R_1+r - \underline{\mathrm{D}}_{1-\alpha} (\rho_{UVB}\| \rho_{UV}\otimes \rho_B)\right) } + v_2^\alpha 2^{\alpha \left( R+R_1 - \underline{I}_{1-\alpha}[V;B|U ]\right)}  \right) \label{BGI}
\end{align}
where $a$ follows \eqref{splitstates}; $b$ follows from \eqref{intermediatetermss}; $c$ follows from \eqref{term11s};  $d$ follows from \eqref{errors2s}, \eqref{errexp1}, \eqref{errexp2} and \eqref{errexp3}.

Thus, 
combining \eqref{splitstates}, \eqref{int11} and \eqref{BGI}, 
we now have the following bound on error probability:
\begin{align}
&\mathbb{E}_{\cC}\Pr \left\{\hat{M} \neq 1 \big| M = 1\right\} \nonumber \\
& \le20\left(v_2^\alpha 2^{\alpha\left(R+R_1+r - \underline{\mathrm{D}}_{1-\alpha} (\rho_{UVB}\| \rho_{UV}\otimes \rho_B)
\right) }+ v_2^\alpha 2^{\alpha \left( R+R_1 - \underline{I}_{1-\alpha}[V:B|U ]\right)}  \right) \nonumber\\
&\hspace{5mm} +{
 \frac{2}{\alpha }\left(\left(\frac{ v_3}{2^{r}}\right)^\alpha
2^{\alpha \underline{\mathrm{D}}_{1+\alpha} (\rho_{US} \| \rho_U \otimes \rho_S) }
+\left(\frac{ v_4}{2^{R_1+r}}\right)^\alpha
2^{\alpha \underline{\mathrm{D}}_{1+\alpha} (\rho_{UVS} \| \rho_{UV} \otimes \rho_S) } \right)   
}.
\end{align}

Therefore, we obtain \eqref{fes} of Lemma \ref{L1}.
In this derivation, 
Lemma \ref{decerr} is used for the evaluation for the first term of \eqref{splitstates}
and 
Lemma \ref{bcove} is used for the evaluation for the second term of \eqref{splitstates}.

\subsection{Secrecy analysis}
Next, we show \eqref{avgsec} of Lemma \ref{L1}.
Our proof employs Lemmas \ref{bcove} and \ref{condcq}, which are given in latter sections.
Let $\rho_{ME}:= \frac{1}{2^{R}} \sum_{m \in [1:2^{R}]} \ket{m}\bra{m}_M \otimes \rho_{E \mid m}$ be the joint state between the register $M$ and $E.$ Notice that if the message $m \in [1:2^{R}]$ is transmitted using the encoding strategy discussed above then the state in Eve's possession at the end of this transmission is the following:
\beq
\label{evestate}
\rho_{E \mid m} = \tr_{B} \left[\cN_{AS \to BE} \left(W^{S' \to ACIJ}_{\cC_{U}, \cC_{m}}\left(\phi_{{S'S}}\right) W^{\dagger S' \to ACIJ}_{\cC_{U}, \cC_{m}} \right)\right].
\enq

We now have the following set of inequalities: 
\begin{align}
&\mathbb{E}_{\cC}\left[ \Pur\left( \rho_{ME} ,  
\tr_{E}[\rho_{ME}] \otimes  \tr_{M}[\rho_{ME}]\right)^2\right] \nonumber\\
=&\mathbb{E}_{\cC}\left[ \Pur\left(\frac{1}{2^R} \sum_{m \in [1:2^R]}\ket{m}\bra{m} \otimes \rho_{E \mid m} , 
\frac{1}{2^R} \sum_{m \in [1:2^R]}\ket{m}\bra{m} \otimes  \sum_{m \in [1:2^R]}\rho_{E}\right)^2\right] \nonumber\\
=   &
\mathbb{E}_{\cC}
\left[
\left(
 \frac{1}{2^R}\sum_{m \in [1:2^R]} \Pur\left(\ket{m}\bra{m} \otimes \rho_{E \mid m} , \ket{m}\bra{m} \otimes  \rho_{E}\right)\right)^2\right] \nonumber\\
\overset{a} \leq  &\frac{1}{2^R}\sum_{m \in [1:2^R]}\mathbb{E}_{\cC}\left[ \Pur\left(\ket{m}\bra{m} \otimes \rho_{E \mid m} , \ket{m}\bra{m} \otimes  \rho_{E}\right)^2\right] \nonumber\\
= &\frac{1}{2^R}\sum_{m \in [1:2^R]}
\mathbb{E}_{\cC}\left[ \Pur\left( \rho_{E \mid m} ,  \rho_{E}\right)^2\right] \nonumber\\
\overset{b} \leq &\frac{2}{2^R}\sum_{m \in [1:2^R]} 
\Big(\mathbb{E}_{\cC} \left[\Pur\left( \rho_{E \mid m},
\frac{1}{2^r}\sum_{i}\rho_{E \mid U(i)}\right)^2\right] 
+ \mathbb{E}_{\cC} \left[\Pur\left( \tr_{M}[\rho_{ME}], \frac{1}{2^r}\sum_{i}\rho_{E \mid U(i)}\right)^2\right] \Big)\nonumber\\
\overset{c} \leq &\frac{4}{2^R}\sum_{m \in [1:2^R]} \mathbb{E}_{\cC} \left[\Pur\left( \rho_{E \mid m},
\frac{1}{2^r}\sum_{i}\rho_{E \mid U(i)}\right)^2\right] \nonumber \\
\overset{d} \leq &\frac{8}{2^R}\sum_{m \in [1:2^R]} \mathbb{E}_{\cC} 
\left[\Pur\left( \rho_{E \mid m}, 
\frac{1}{2^{(R_1+r)}} \sum_{(i,j)} \rho_{E \mid U(i), V(m,i,j)}\right)^2\right] 
\nonumber \\
&
+ \frac{8}{2^R}\sum_{m \in [1:2^R]} \mathbb{E}_{\cC} \left[\Pur\left(
\frac{1}{2^r}\sum_{i}\rho_{E \mid U(i)}, 
\frac{1}{2^{(R_1+r)}} \sum_{(i,j)} \rho_{E \mid U(i), V(m,i,j)}
\right)^2\right] \nonumber \\
\label{secrecypurbound}
\overset{e}  \leq & \frac{8}{2^R}\sum_{m \in [1:2^R]} 
\mathbb{E}_{\cC}
\left[
\Pur\left( \rho_{E \mid m}, 
\frac{1}{2^{(R_1+r)}} \sum_{(i,j)} \rho_{E \mid U(i), V(m,i,j)}\right)^2
\right] 
\nonumber \\
&+ \frac{8}{2^{R}}
\sum_{m \in [1:2^R]} \frac{1}{2^r}\sum_{i}
\mathbb{E}_{\cC} \left[\Pur\left( \rho_{E \mid U(i)},
\frac{1}{2^{R_1}} \sum_{j} \rho_{E \mid U(i), V(m,i,j)}
\right)^2\right] 
\end{align}
where
$a$ follows from the convexity of $x \mapsto x^2$;
$b$ follows from Fact \ref{fact:trianglepurified} (the triangle inequality for the purified distance) 
and the relation that $\tr_{M}[\rho_{ME}]=\rho_E$;
$c$ follows from the inequality
$\Pur\left( \tr_{M}[\rho_{ME}],
\frac{1}{2^r}\sum_{i}\rho_{E \mid U(i)}\right)
\le \frac{1}{2^R}\sum_{m \in [1:2^R]} \mathbb{E}_{\cC} \left[\Pur\left( \rho_{E \mid m},
\frac{1}{2^r}\sum_{i}\rho_{E \mid U(i)}\right)\right]$, which can be shown by the relation
$\frac{1}{2^R}\sum_{m \in [1:2^R]} \rho_{E \mid m}=\tr_{M}[\rho_{ME}]$;
$d$ follows from Fact \ref{fact:trianglepurified};
$e$ 
follows from Fact \ref{fact:monotonequantumoperation} and convexity of square function;

We now bound each of the terms on the right hand side of \eqref{secrecypurbound}. Towards this consider the first term:
\begin{align}
&
 \mathbb{E}_{\cC}\bigg[
\Pur\left(
\frac{1}{2^{(R_1+r)}} \sum_{(i,j)} \rho_{E \mid U(i), V(m,i,j)},
 \rho_{E \mid m}
 \right)^2 \bigg] \nonumber\\
\overset{a}=& 
 \mathbb{E}_{\cC}\bigg[\Pur\bigg(\frac{1}{2^{(R_1+r)}} \sum_{(i,j) \in [1:2^{r}] \times [1:2^{R_1}]} 
\tr_{CB} \cN_{AS\to BE } \left(\rho_{CAS \mid U(i)V(m,i,j)} \right), \nonumber\\
&\hspace{15mm}\tr_{CBIJ} \cN_{AS\to BE } \left(W^{S' \to ACIJ}_{\cC_{U}, \cC_{m}}\left(\phi_{{S'S}}\right) W^{\dagger S' \to ACIJ}_{\cC_{U}, \cC_{m}} \right)\bigg)^2\bigg] \nonumber\\
\overset{b}\le &
 \frac{1}{\alpha }\left(\left(\frac{ v_3}{2^{r}}\right)^\alpha
2^{\alpha \underline{\mathrm{D}}_{1+\alpha} (\rho_{US} \| \rho_U \otimes \rho_S) }
+\left(\frac{ v_4}{2^{R_1+r}}\right)^\alpha
2^{\alpha \underline{\mathrm{D}}_{1+\alpha} (\rho_{UVS} \| \rho_{UV} \otimes \rho_S) } \right)  ,
\label{sec1}
\end{align}
where
$a$ follows from \eqref{evestate};
$b$ can be shown by replacing $B$ by $E$ in the derivation of \eqref{int11}.


We now bound the second term in \eqref{secrecypurbound} as follows: 
\begin{align}
 & \mathbb{E}_{\cC} \left[\Pur\left( \rho_{E \mid U(i)}, 
 \frac{1}{2^{R_1}} \sum_{j} \rho_{E \mid U(i), V(m,i,j)}\right)^2\right] \nonumber \\
 &\overset{a} \leq  \mathbb{E}_{\cC} 
 \Big[1 - 2^{\left(\underline{\mathrm{D}}_{1+\alpha}\left(   \frac{1}{2^{R_1}} \sum_{j} \rho_{E \mid U(i), V(m,i,j)} \|  \rho_{E \mid U(i)}\right) \right)}\Big] \nonumber\\ 
 &\overset{c} \leq 
{\ln 2 \cdot \mathbb{E}_{\cC}\left[\underline{\mathrm{D}}_{1+\alpha}\left(   \frac{1}{2^{R_1}} \sum_{j} \rho_{E \mid U(i), V(m,i,j)} \|  \rho_{E \mid U(i)}\right) \right]}
  \nonumber 
\\
 &\overset{d} \leq {
 \frac{1}{\alpha }\left(\frac{v_5^\alpha}{2^{\alpha R_1}} 2^{\alpha{\underline{\mathrm{ D}}}_{1+ \alpha} \left(\rho_{UVE} \| \rho_{U-V-E} \right)}\right) 
 },
 \label{sec2}
\end{align}
where $a$ follows from Fact \ref{pinsker}; $b$ follows from the Fact \ref{jenssen}; 
$c$ follows from \eqref{EH1}; and  
$d$ follows from Lemma \ref{condcq}.

Thus, from \eqref{secrecypurbound}, \eqref{sec1} and \eqref{sec2} we have the following bound: 

\begin{align}
&\mathbb{E}_{\cC}\left[ \Pur\left( \rho_{ME} ,  \tr_{E}[\rho_{ME}] \otimes  \tr_{M}[\rho_{ME}]\right)^2
\right] \nonumber\\
&\le {
\frac{8}{\alpha }\left(\left(\frac{ v_3}{2^{r}}\right)^\alpha
2^{\alpha \underline{\mathrm{D}}_{1+\alpha} (\rho_{US} \| \rho_U \otimes \rho_S) }
+\left(\frac{ v_4}{2^{R_1+r}}\right)^\alpha
2^{\alpha \underline{\mathrm{D}}_{1+\alpha} (\rho_{UVS} \| \rho_{UV} \otimes \rho_S) } \right)  
} \nonumber\\
&\hspace{5mm}+ { 
\frac{8}{\alpha }\left(\frac{v_5^\alpha}{2^{\alpha R_1}} 2^{\alpha{\underline{\mathrm{ D}}}_{1+ \alpha} \left(\rho_{UVE} \| \rho_{U-V-E} \right)}\right)    
},
\end{align}
which implies \eqref{avgsec}. In this derivation, 
Lemma \ref{bcove} is used for the evaluation of the first term in \eqref{secrecypurbound}
and 
Lemma \ref{condcq} is used for the evaluation of the second term in \eqref{secrecypurbound}.

\section{Hypothesis testing}\label{pinchsec}
We consider hypothesis testing on three quantum systems $U,V$, and $B$.
The null hypothesis is $\rho_{UVB}$, and the alternative hypothesis is composed of
the product state $ \rho_{UV}\otimes \rho_B$ and the state $\rho_{V-U-B} $.
To give our test, we fix two real numbers $\sM_1$ and $\sM_2$ and define the following projectors:
\begin{align}\label{puv}
\Pi_1:=&\{ {\cal E}_2(\rho_{UVB}) \ge \sM_1 \rho_{UV}\otimes \rho_B\},\\
\label{pvu}\Pi_2:=&\{ {\cal E}_2(\rho_{UVB}) \ge \sM_2 {\cal E}_1(\rho_{V-U-B})\}.
\end{align}
Since it follows from the property of the pinching operations defined above that $\Pi_1$ and $\Pi_2$ commute,
the test $\Pi:=\Pi_1\Pi_2$ satisfies the properties 
$\Pi \leq \Pi_1, \Pi_2$ and $\left(\mathbb{I}-\Pi\right) \leq (\mathbb{I}-\Pi_1)+(\mathbb{I}-\Pi_2).$
The following lemma shows the performance of the test $\Pi$.
\begin{lemma}
\label{decerr}
For $\alpha \in (0,1)$
\begin{align}
\label{ineq1}
\Tr \Pi_1 (\rho_{UV}\otimes \rho_B )
& \le v_2^\alpha
 \sM_1^{-(1-\alpha)} 
2^{-s \underline{\mathrm{D}}_{1-\alpha} (\rho_{UVB}\| \rho_{UV}\otimes \rho_B) };
\\
\label{ineq2}
\Tr \Pi_2 \rho_{U-V-B} 
&\le v_2^\alpha \sM_2^{-(1-\alpha)} 2^{-\alpha 
\underline{I}_{1-\alpha}[V;B|U ]_{\rho_{UVB}|\rho_{UV}}};
\\
\label{ineq3}
\Tr (\mathbb{I}-\Pi_1) \rho_{UVB}
& \le v_2^\alpha \sM_1^{\alpha} 
2^{-\alpha \underline{\mathrm{D}}_{1-\alpha} (\rho_{UVB}\| \rho_{UV}\otimes \rho_B) };
\\
\label{ineq4}
\Tr (\mathbb{I}-\Pi_2) \rho_{UVB}
& \le v_2^\alpha \sM_2^{\alpha} 
2^{-\alpha \underline{I}_{1-\alpha}[V;B|U ]_{\rho_{UVB}|\rho_{UV}}}.
\end{align}
\end{lemma}

\begin{proof}
We will only give the proof for \eqref{ineq1}. The proof for the other inequalities follows using exactly similar techniques.

Notice the following set of inequalities. 
\begin{align}
 &\Tr\left[ (\mathbb{I}-\Pi_1) \rho_{UVB} \right] \nonumber \\
 \overset{a}=&  \Tr \left[\cE_2(\mathbb{I}-\Pi_1)\rho_{UVB}\right] \nonumber\\
\overset{b} = &   \Tr\left[ (\mathbb{I}-\Pi_1) \cE_2(\rho_{UVB})\right] \nonumber\\
= &  \Tr\left[ (\mathbb{I}-\Pi_1) \left(\cE_2(\rho_{UVB})\right)^{1-\alpha} \left(\cE_2(\rho_{UVB})\right)^{\alpha}\right] \nonumber\\
\overset{c}\le&  \sM_1^{\alpha}
\Tr \left[(\mathbb{I}-\Pi_1) \cE_2 (\rho_{UVB})^{1-\alpha}
(\rho_{UV}\otimes \rho_S)^{\alpha}\right] \nonumber\\
\overset{d}\le&  
\sM_1^{\alpha}
\Tr  \left[\cE_2 (\rho_{UVB})^{1-\alpha}
(\rho_{UV}\otimes \rho_S)^{\alpha}\right] \nonumber\\
= & \sM_1^{\alpha }
\Tr \bigg[
\Big( 
(\rho_{UV}\otimes \rho_B)^{\frac{\alpha}{2(1-\alpha)}}
 \cE_2 (\rho_{UVB})
(\rho_{UV}\otimes \rho_B)^{\frac{\alpha}{2(1-\alpha)}}
\Big)^{1-\alpha} \bigg] \nonumber\\
= & \sM_1^{\alpha }
\Tr \Big[
\Big( 
(\rho_{UV}\otimes \rho_B)^{\frac{\alpha}{2(1-\alpha)}}
 \cE_2 (\rho_{UVB})
(\rho_{UV}\otimes \rho_B)^{\frac{\alpha}{2(1-\alpha)}}
\Big) \cdot\Big( 
(\rho_{UV}\otimes \rho_B)^{\frac{\alpha}{2(1-\alpha)}}
 \cE_2 (\rho_{UVB})
(\rho_{UV}\otimes \rho_B)^{\frac{\alpha}{2(1-\alpha)}}
\Big)^{-\alpha} \Big] \nonumber\\
=
 & \sM_1^{\alpha }
\Tr \Big[
\Big( 
(\rho_{UV}\otimes \rho_B)^{\frac{\alpha}{2(1-\alpha)}}
\rho_{UVB}
(\rho_{UV}\otimes \rho_B)^{\frac{\alpha}{2(1-\alpha)}}
\Big)\cdot
\Big( 
(\rho_{UV}\otimes \rho_B)^{\frac{\alpha}{2(1-\alpha)}}
\cE_2(\rho_{UVB})
(\rho_{UV}\otimes \rho_B)^{\frac{\alpha}{2(1-\alpha)}}
\Big)^{-\alpha} \Big] \nonumber\\
\stackrel{e}{\le}
& v_2^{\alpha}\sM_1^{\alpha }
\Tr \Big[
\Big( 
(\rho_{UV}\otimes \rho_B)^{\frac{\alpha}{2(1-\alpha)}}
\rho_{UVB}
(\rho_{UV}\otimes \rho_B)^{\frac{\alpha}{2(1-\alpha)}}
\Big) \cdot
\Big( 
(\rho_{UV}\otimes \rho_B)^{\frac{\alpha}{2(1-\alpha)}}
 (\rho_{UVB})
(\rho_{UV}\otimes \rho_B)^{\frac{\alpha}{2(1-\alpha)}}
\Big)^{-\alpha} \Big] \nonumber\\
= & v_2^{\alpha}\sM_1^{\alpha }
\Tr 
\Big( 
(\rho_{UV}\otimes \rho_B)^{\frac{\alpha}{2(1-\alpha)}}
\rho_{UVB}
(\rho_{UV}\otimes \rho_B)^{\frac{\alpha}{2(1-\alpha)}}
\Big)^{1-\alpha} \nonumber\\
= & v_2^{\alpha}\sM_1^{\alpha } 
2^{-\alpha \underline{\mathrm{D}}_{1-\alpha}(\rho_{UVB} \|\rho_{UV}\otimes \rho_B)} ,
\label{ER1}
\end{align}
where $a$ and $b$ both follow from the definition of $\Pi_1$ and $\cE_2$ along with the fact that after applying the pinching operation $\cE_2,$ $(\mathbb{I}-\Pi_1)$ and $\cE_2(\rho_{UVB})$ commute; $c$ follows from the definition of $\Pi_1;$ $d$ follows from the monotonicity of the trace operation and $e$ follows because of the Fact \ref{pinchs} and the operator monotonicity of the function $x^{-\alpha}.$
\end{proof}

Here, we discuss the relation between Lemma \ref{decerr}
and existing results for quantum hypothesis testing.
For this aim, 
we consider two states $\rho$ and $\sigma$ on the single system
because since existing studies mainly discuss 
such a case.
Let ${\cal E}_{\sigma}$ be the pinching with respect to 
\begin{align}
\Pi_\sigma:= \{ {\cal E}_{\sigma}(\rho)\ge \sM \sigma \}.
\end{align}
Applying the same method as the proof of Lemma \ref{decerr},
we have
\begin{align}
\label{ineq1X}
\begin{aligned}
\Tr \Pi_\sigma \sigma
& \le v_\sigma^\alpha
\sM^{-(1-\alpha)} 
2^{-s \underline{\mathrm{D}}_{1-\alpha} (\rho\| \sigma) };
\\
\Tr (\mathbb{I}-\Pi_\sigma) \rho
& \le v_\sigma^\alpha \sM^\alpha 
2^{-\alpha \underline{\mathrm{D}}_{1-\alpha} (\rho\| \sigma) },
\end{aligned}
\end{align}
where $v_\sigma$ is the number of distinct eigenvalues of $\sigma$.
In contrast, 
the paper \cite{Hoeffding}
showed
\begin{align}
\label{ineq1Y}
\begin{aligned}
\Tr \Pi \sigma
& \le 
\sM^{-(1-\alpha)} 
2^{-s {\mathrm{D}}_{1-\alpha} (\rho\| \sigma) };
\\
\Tr (\mathbb{I}-\Pi) \rho
& \le \sM^\alpha 
2^{-\alpha {\mathrm{D}}_{1-\alpha} (\rho\| \sigma) },
\end{aligned}
\end{align}
with
$\Pi:= \{ \rho\ge \sM \sigma \}$
and $-\alpha {\mathrm{D}}_{1-\alpha} (\rho\| \sigma)
:= \log \Tr \rho^{1-\alpha}\sigma^{\alpha}$.

Since ${\mathrm{D}}_{1-\alpha} (\rho\| \sigma)
\ge  \underline{\mathrm{D}}_{1-\alpha} (\rho\| \sigma)$,
the evaluation \eqref{ineq1Y}
is better than the evaluation \eqref{ineq1X}.
While the evaluation \eqref{ineq1Y} is obtained from the optimal testing
$\Pi$, whose optimality is shown in \cite{Helstrom,Holevo}\cite[Eq. (14)]{Nagaoka},
the evaluation \eqref{ineq1X} is obtained from the testing
$\Pi_\sigma$ based on the pinching ${\cal E}_{\sigma}$, which is not optimal in general.
Hence, to address the merit of the evaluation \eqref{ineq1X},
we compare it with existing evaluation for the error probability of the same testing $\Pi_\sigma$.
Before the paper \cite{Hoeffding},
the paper \cite{Ogawa} showed the following evaluation;
\begin{align}
\label{ineq1Z}
\begin{aligned}
\Tr \Pi_\sigma \sigma
& \le v_\sigma^\alpha
\sM^{-(1-\alpha)} 
2^{-s \hat{\mathrm{D}}_{1-\alpha} (\rho\| \sigma) };
\\
\Tr (\mathbb{I}-\Pi_\sigma) \rho
& \le v_\sigma^\alpha \sM^\alpha 
2^{-\alpha \hat{\mathrm{D}}_{1-\alpha} (\rho\| \sigma) },
\end{aligned}
\end{align}
where $-\alpha \hat{\mathrm{D}}_{1-\alpha} (\rho\| \sigma)
:= \log \Tr \rho \sigma^{\alpha/2}\rho^{-\alpha}\sigma^{\alpha/2}$.
One might consider that we can replace Lemma \ref{decerr}
by the evaluation similar to \eqref{ineq1Z}.
However, the information processing inequality of
$\hat{\mathrm{D}}_{1-\alpha} (\rho\| \sigma)$ has not been shown.
Since we employ the information processing inequality of
$\underline{\mathrm{D}}_{1-\alpha} (\rho\| \sigma)$ for $\alpha \in (0,1/2]$ 
in the latter discussion,
we cannot replace Lemma \ref{decerr}
by such an evaluation.

\section {Bivariate classical-quantum channel resolvability lemma with error exponent}\label{s8}  

\begin{lemma}
\label{bcove}
Let $\rho_{UVS}:= \sum_{(u,v) \in  \mathcal{U} \times \mathcal{V}} p{(u,v)}(u,v)\ket{u}\bra{u}_U \otimes \ket{v}\bra{v}_V \otimes \rho_{S \mid u,v}$ be a classical-quantum state. Let,
 $\big\{U(1),$ $ \cdots, U({2^{r}})\big\},$ be a set of independent and identically distributed random variables where for every $ i \in [1:2^{r}],$ $U_{i} \sim p_{U}.$ Further for every $(i, j) \in [1:2^{r}]\times[1:2^{R}],$ let  $\left\{V{(1,1)},\cdots, V{(2^{r},2^{R})}\right\}$ be a collection of independent sequences and for every $(i,j),$ $V{(i,j)}\sim p_{V \mid U (i)} $ Let  $\cC:= \left\{U(1)\cdots  U(2^{r}), V(1,1) \cdots V{(2^{r},2^{R})}\right\}$ and $\tau_{S \mid \cC}  := \frac{1}{2^{(R+r)}} \sum_{(i,j)}\rho_{S \mid U{(i)},V{(i,j)}}.$
Then for $\alpha \in [0,1]$ there exists constants $v_3,v_4> 0$ such that, 
\begin{align*}
\mathbb{E}_{\cC}\left[\underline{\mathrm{D}}_{1+\alpha}\left(\tau_{S \mid \cC} \| \rho_S\right) \right]
\le &\frac{1}{\alpha} \log_2 \left(\rE_{\cC}\left[2^{\alpha\underline{\mathrm{D}}_{1+\alpha}\left(\tau_{S \mid \cC} \| \rho_S\right) }\right]\right)\\
\leq& \frac{1}{ \alpha\ln 2}\left(\left(\frac{ v_3}{2^{r}}\right)^\alpha
2^{\alpha \underline{\mathrm{D}}_{1+\alpha} (\rho_{US} \| \rho_U \otimes \rho_S) }
+
\left(\frac{ v_4}{2^{R+r}}\right)^\alpha
2^{\alpha \underline{\mathrm{D}}_{1+\alpha} (\rho_{UVS} \| \rho_{UV} \otimes \rho_S) } \right),
\end{align*}
where in the above the first inequality follows because of the concavity of the $\log (\cdot).$
\end{lemma}
\begin{proof}
In the proof of this lemma we will need a pair of pinching maps similar to the pair $\cE_1$ and $\cE_2$ defined in earlier sections. Towards this consider the following states: 
\begin{align*}
\rho_{UVS} &=   \sum_{(u,v) \in  \mathcal{U} \times \mathcal{V}} p{(u,v)}\ket{u}\bra{u}_U \otimes \ket{v}\bra{v}_V \otimes \rho_{S \mid u,v};\\
 \rho_{V-U-S} & = \sum_{u\in  \mathcal{U}} p_U{(u)}
 \ket{u}\bra{u}_U \otimes\rho_{V \mid u} \otimes \rho_{S \mid u},
\end{align*}
where in the above $\rho_{V \mid u}$ and $\rho_{S \mid u}$ are appropriate marginals of the state 
$\rho_{UVS}$, and $p_U(u)$ is the marginal distribution with respect to $U$.
Here, the pinching operations $\cE_3$ and  $\cE_4$ are defined in Section \ref{SS4}.
Notice that $\cE_3$ and  $\cE_4$  are defined in a manner similar to $\cE_1$ and  $\cE_2$, by replacing the system $B$ with the system $S$. 

We now have the following set of inequalities: 
\begin{align*}
&\rE_{\cC}\left[2^{\alpha\underline{\mathrm{D}}_{1+\alpha}\left(\tau_{S \mid \cC} \| \rho_S\right) }\right]\\
=&
\rE_{\cC} 
\Tr \left[
\left(\rho_S^{-\frac{\alpha}{2(1+\alpha)}}
\tau_{S \mid \cC}
\rho_S^{-\frac{\alpha}{2(1+\alpha)}}
\right)^{1+\alpha}\right]\\
=&
\rE_{\cC} 
\Tr \left[
\left(\rho_S^{-\frac{\alpha}{2(1+\alpha)}}
\frac{1}{2^{(R+r)}} \sum_{(i,j)}\rho_{S \mid U{(i)},V{(i,j)}}
\rho_S^{-\frac{\alpha}{2(1+\alpha)}}
\right)^{1+\alpha}\right]\\
=&
\frac{1}{2^{(R+r)}} \sum_{(i,j)}\rE_{\cC} 
\Tr \left[
\left(\rho_S^{-\frac{\alpha}{2(1+\alpha)}}
\rho_{S \mid U{(i)},V{(i,j)}}
\rho_S^{-\frac{\alpha}{2(1+\alpha)}}\right)
\cdot
\left(
\rho_S^{-\frac{\alpha}{2(1+\alpha)}}\frac{1}{2^{(R+r)}} \sum_{(i',j')}\rho_{S \mid U{(i')},V{(i',j')}}
\rho_S^{-\frac{\alpha}{2(1+\alpha)}}
\right)^{\alpha}\right]\\
=&\frac{1}{2^{(R+r)}} \sum_{(i,j)}
\rE_{\cC} 
\Tr 
\bigg[\left(
\rho_S^{-\frac{\alpha}{2(1+\alpha)}}
\rho_{S \mid U{(i)},V{(i,j)}}
\rho_S^{-\frac{\alpha}{2(1+\alpha)}}
\right) \\
&\cdot \left(
\rho_S^{-\frac{\alpha}{2(1+\alpha)}}
\frac{1}{2^{R+r}}
\left(\rho_{S \mid U{(i)},V{(i,j)}}
+
\sum_{ j'\neq j}
\rho_{S \mid U{(i)},V{(i,j')}}
+
\sum_{i' \neq i, j' \neq j}
\rho_{S \mid U{(i')},V{(i',j')}}
\right)
\rho_S^{-\frac{\alpha}{2(1+\alpha)}}
\right)^{\alpha}\bigg]\\
\overset{a} \le &\frac{1}{2^{(R+r)}} \sum_{(i,j)}
\rE_{\cC} 
\Tr 
\bigg[\left(
\rho_S^{-\frac{\alpha}{2(1+\alpha)}}
\rho_{S \mid U{(i)},V{(i,j)}}
\rho_S^{-\frac{\alpha}{2(1+\alpha)}}
\right) \\
&\cdot \left(
\rho_S^{-\frac{\alpha}{2(1+\alpha)}}
\frac{1}{2^{R+r}}
\left(
\rho_{S \mid U{(i)},V{(i,j)}}
+
\sum_{j' \neq j}\rE_{V \mid U}
\left[\rho_{S \mid U{(i)},V{(i,j')}}\right]
+
\sum_{i' \neq i, j' \neq j}
\rE_{U,V}\left[\rho_{S \mid U{(i')},V{(i',j')}}\right]\right)
\rho_S^{-\frac{\alpha}{2(1+\alpha)}}
\right)^{\alpha}\bigg]\\
\overset{b}  \le &\frac{1}{2^{(R+r)}} \sum_{(i,j)}
\rE_{\cC} 
\Tr 
\bigg[\left(
\rho_S^{-\frac{\alpha}{2(1+\alpha)}}
\rho_{S \mid U{(i)},V{(i,j)}}
\rho_S^{-\frac{\alpha}{2(1+\alpha)}}
\right) \cdot \left(
\rho_S^{-\frac{\alpha}{2(1+\alpha)}}
\frac{1}{2^{R+r}}
\left(
\rho_{S \mid U{(i)},V{(i,j)}}
+
2^R \rho_{S \mid U(i)}
+
2^{R+r}
\rho_S\right)
\rho_S^{-\frac{\alpha}{2(1+\alpha)}}
\right)^{\alpha}\bigg]\\
\overset{c}\le &
 \sum_{i,j}\frac{1}{2^{R+r}}
\rE_{U(i),V(i,j)} 
\Tr 
\bigg[\left(
\rho_S^{-\frac{\alpha}{2(1+\alpha)}}
\rho_{S \mid U{(i)},V{(i,j)}}
\rho_S^{-\frac{\alpha}{2(1+\alpha)}}
\right) \\
&\cdot \left(
\rho_S^{-\frac{\alpha}{2(1+\alpha)}}
\frac{1}{2^{R+r}}
\left(
v_4 \cE_{4|U(i)} (\rho_{S \mid U(i),V(i,j)})
+
v_3
2^R
\cE_3(\rho_{S|U(i)})
+
2^{R+r}
\rho_S\right)
\rho_S^{-\frac{\alpha}{2(1+\alpha)}}\right)^\alpha
\bigg]\\
\overset{d}\le&\sum_{i,j}\frac{1}{2^{R+r}}
\rE_{U(i),V(i,j)} 
\Tr 
\bigg[\left(
\rho_S^{-\frac{\alpha}{2(1+\alpha)}} \rho_{S \mid U(i),V(i,j)}\rho_S^{-\frac{\alpha}{2(1+\alpha)}}
\right) \\
&\cdot \left(
\rho_S^{-\frac{\alpha^2}{2(1+\alpha)}}
\frac{1}{2^{\alpha(R+r)}}
\left(
v_4^\alpha (\cE_{4|U(i)} (\rho_{S \mid U(i),V(i,j)}))^\alpha
+
v_3^\alpha
2^{\alpha R}
(\cE_3(\rho_{S|U(i)}))^\alpha
+
2^{\alpha(R+r)}
\rho_S^\alpha\right)
\rho_S^{-\frac{\alpha^2}{2(1+\alpha)}}\right)
\bigg]\\
\overset{e}=& 1 + \rE_{U',V'}
\tr \left[ \frac{v_4^\alpha}{2^{\alpha (r+R)}} \rho_{S \mid U',V'} 
(\cE_{4|U'} (\rho_{S \mid U',V'}))^\alpha \rho_S^{-\alpha}\right] 
+ \rE_{U',V'}\tr \left[\frac{v_3 ^ \alpha}{2^{\alpha r}} \rho_{S \mid U',V'}\cE_3(\rho_{S|U'}))^\alpha
\rho_S^{-\alpha} \right] \\
\overset{f}= & 1 +  \frac{v_4^{\alpha}}{2^{\alpha (r+R)}} \tr\left[\rE_{U',V'}(\cE_{4|U'} 
(\rho_{S \mid U',V'}))^{(1+\alpha)} \rho_S^{-\alpha} \right] 
+ \frac{v_3^{\alpha}}{2^{\alpha r}} 
\tr\left[\rE_{U'}(\cE_{3} (\rho_{S \mid U'}))^{(1+\alpha)} \rho_S^{-\alpha} \right]\\
\overset{g}\le& 1 + \frac{v_3 ^ {\alpha}}{2^{\alpha r}} 2^{\alpha\mathrm{\underline{D}}_{1+\mathrm{\alpha}}\left(\rho_{US} \| \rho_U \otimes \rho_S\right)} + \frac{v_4 ^ {\alpha}}{2^{\alpha (r+R)}} 2^{\alpha\mathrm{\underline{D}}_{1+\mathrm{\alpha}}\left(\rho_{UVS} \| \rho_{UV} \otimes \rho_S\right)},
\end{align*}
where from step $e$, the variables $U'$ and $V'$ are subject to the joint distribution $p(u,v)$.
Here, $a$ follows from the fact that when $j \neq j'$ then the random variables $(V(i,j),V(i,j'))$ are independent of each other and from the operator Jensen's inequality; 
$b$ follows from symmetry and from the definition of $\rho_{S \mid u}$ and $\rho_S$; 
$c$ follows from the Fact \ref{pinchs}; $d$ follows because the terms in the second terms inside the trace commute and from the fact that $(a+b)^x \leq a^x+b^x (a,b>0; x<1);$ $e$ follows from the circular and linear property of the trace operation; $f$ follows from the circularity of trace operation and 
$g$ follows from the definition of $\mathrm{\underline{D}}_{1+\alpha}(\cdot \| \cdot)$,
the data-processing inequality (Fact \ref{fact:monotonequantumoperation}),
and the fact that the states involved are classical-quantum states. 
The desired bound now follows from the fact that $\log_2(1+x) \leq \frac{x}{\ln2}.$ 
\end{proof}

\section {Conditional classical-quantum channel resolvability lemma with error exponent }\label{s9}
\begin{lemma}
\label{condcq}
Let $\rho^{UVE}:= \sum_{(u,v) \in  \mathcal{U} \times \mathcal{V}} p_{UV}{(u,v)}(u,v)\ket{u}\bra{u}^U \otimes \ket{v}\bra{v}^V \otimes \rho_{E \mid u,v}$ be a classical-quantum state. Further, let 
$\cC:= \left\{U', V(1), \cdots, V(2^{R})\right\}$ be a collection of  random variables where for  every $i \in [1:2^{R}],$ 
$(U',V(i)) \sim p_{UV}$ and for 
$i \neq i',$ $(V(i),V(i')) \sim p_{V(i) \mid U'}\cdot p_{V(i') \mid U'}.$ 
Consider the following state:
\begin{align*}
\tau_{E \mid \cC} & := \frac{1}{2^{R}} \sum_{i}\rho_{E \mid U',V(i)} .
\end{align*}
Then, for $\alpha \in [0,1]$ there exists constant $v_5> 0$ such that, 
\begin{align*}
\mathbb{E}_{\cC}\left[\underline{\mathrm{D}}_{1+\alpha}\left(\tau_{E \mid \cC} \| \rho_{E \mid U'}\right) \right]
\le &\frac{1}{\alpha} \log_2\left(\rE_{\cC}\left[2^{\alpha\underline{\mathrm{D}}_{1+\alpha}\left(\tau_{E \mid \cC} \| \rho_{E \mid U'}\right) }\right]\right)\\
\le&\frac{1}{\alpha\ln 2}\left(\frac{v_5^\alpha}{2^{\alpha R}} 2^{\alpha \underline{\mathrm{D}}_{1+ \alpha} \left(\rho_{UVE} \| \rho_{V-U-E} \right)}\right),
\end{align*}
where in the above $\rho_{V-U-E}:= \sum_{u}p_{U}(u)\ket{u}\bra{u}_U\otimes\rho_{V \mid u} \otimes \rho_{E \mid u}$ and the first inequality is because of the concavity of the $\log (\cdot).$
\end{lemma}
\begin{proof}
Let $\cE_{5 \mid u}$ be the pinching maps with respect to the spectral decomposition of $\rho_{E \mid u}.$  Further, let $v_5$ represent  
the maximum number of distinct components of the pinching map
$\{{\cal E}_{5 \mid u}\}_{u}.$  

We now have the following inequalities: 
\begin{align*}
&\rE_{\cC}\left[2^{\alpha\underline{\mathrm{D}}_{1+\alpha}\left(\tau_{E \mid \cC} 
\| \rho_{E \mid U'}\right) }\right]\\
 =&
\rE_{\cC} 
\Tr 
\left[\left(
\rho_{E \mid U'}^{-\frac{\alpha}{2(1+\alpha)}}
\tau_{E \mid \cC}
\rho_{E \mid U'}^{-\frac{\alpha}{2(1+\alpha)}}
\right)^{1+\alpha}\right]\\
=&
\rE_{\cC} 
\Tr \left[\left(
\rho_{E \mid U'}^{-\frac{\alpha}{2(1+\alpha)}}
\frac{1}{2^{R}} \sum_{i}\rho_{E \mid U',V{(i)}}
\rho_{E \mid U'}^{-\frac{\alpha}{2(1+\alpha)}}
\right)^{1+\alpha}\right]\\
=&
\frac{1}{2^{R}} \sum_{i}\rE_{\cC} 
\Tr \left[
\left(\rho_{E \mid U'}^{-\frac{\alpha}{2(1+\alpha)}}
\rho_{E \mid U',V(i)}
\rho_{E \mid U'}^{-\frac{\alpha}{2(1+\alpha)}}\right)
\cdot
\left(
\rho_{E \mid U' }^{-\frac{\alpha}{2(1+\alpha)}}\frac{1}{2^{R}} \sum_{i'}
\rho_{E \mid U',V(i')}
\rho_{E \mid U'}^{-\frac{\alpha}{2(1+\alpha)}}
\right)^{\alpha}\right]\\
=&\frac{1}{2^{R}} \sum_{i}\rE_{\cC} \Tr \left[
\left(\rho_{E \mid U'}^{-\frac{\alpha}{2(1+\alpha)}}
\rho_{E \mid U',V(i)}
\rho_{E \mid U'}^{-\frac{\alpha}{2(1+\alpha)}}\right)
\cdot
\left(
\rho_{E \mid U' }^{-\frac{\alpha}{2(1+\alpha)}}\frac{1}{2^{R}} 
\left(\rho_{E \mid U',V(i)} + \sum_{i' \neq i}\rho_{E \mid U',V(i')}\right)
\rho_{E \mid U'}^{-\frac{\alpha}{2(1+\alpha)}}
\right)^{\alpha}\right]\\
\overset{a} \le&\frac{1}{2^{R}} \sum_{i}\rE_{\cC} \Tr \left[
\left(\rho_{E \mid U'}^{-\frac{\alpha}{2(1+\alpha)}}
\rho_{E \mid U',V(i)}
\rho_{E \mid U'}^{-\frac{\alpha}{2(1+\alpha)}}\right)
\cdot
\left(
\rho_{E \mid U' }^{-\frac{\alpha}{2(1+\alpha)}}\frac{1}{2^{R}} 
\left(\rho_{E \mid U',V(i)} + 2^R\rho_{E \mid U'}\right)
\rho_{E \mid U'}^{-\frac{\alpha}{2(1+\alpha)}}
\right)^{\alpha}\right]\\
\overset{b}\le&\frac{1}{2^{R}} \sum_{i}\rE_{\cC} \Tr \left[
\left(\rho_{E \mid U'}^{-\frac{\alpha}{2(1+\alpha)}}
\rho_{E \mid U',V(i)}
\rho_{E \mid U'}^{-\frac{\alpha}{2(1+\alpha)}}\right)
\cdot
\left(
\rho_{E \mid U' }^{-\frac{\alpha}{2(1+\alpha)}}\frac{1}{2^{R}} 
\left(v_5\cE_{5 \mid U'}\left(\rho_{E \mid U,V(i)}\right) + 
2^R\rho_{E \mid U'}\right)
\rho_{E \mid U'}^{-\frac{\alpha}{2(1+\alpha)}}
\right)^{\alpha}\right]\\
\overset{c}=&\rE_{U',V'} \Tr \left[
\left(\rho_{E \mid U'}^{-\frac{\alpha}{2(1+\alpha)}}
\rho_{E \mid U',V'}
\rho_{E \mid U'}^{-\frac{\alpha}{2(1+\alpha)}}\right)
\cdot
\left(
\rho_{E \mid U' }^{-\frac{\alpha}{2(1+\alpha)}}\frac{1}{2^{R}} 
\left(v_5\cE_{5 \mid U'}\left(\rho_{E \mid U',V'}\right) + 2^{R}\rho_{E \mid U'}\right)
\rho_{E \mid U'}^{-\frac{\alpha}{2(1+\alpha)}}
\right)^{\alpha}\right]\\
\overset{d} \le &\rE_{U',V'} \Tr \left[
\left(\rho_{E \mid U'}^{-\frac{\alpha}{2(1+\alpha)}}
\rho_{E \mid U',V'}
\rho_{E \mid U'}^{-\frac{\alpha}{2(1+\alpha)}}\right)
\cdot
\left(
\rho_{E \mid U' }^{-\frac{\alpha^2}{2(1+\alpha)}}\frac{1}{2^{\alpha R}} 
\left(v_5^\alpha\left(\cE_{5 \mid U'}\left(\rho_{E \mid U',V'}\right)\right)^\alpha 
+ 2^{\alpha R}\rho_{E \mid U'}^\alpha\right)
\rho_{E \mid U'}^{-\frac{\alpha^2}{2(1+\alpha)}}
\right)\right]\\
\overset{e}=& 1 + \rE_{U',V'}\tr \left[ \frac{v_5^\alpha}{2^{\alpha R}} 
\rho_{E \mid U',V'} (\cE_{5 \mid U'} (\rho_{E \mid U',V'}))^\alpha 
\rho_{E \mid U'}^{-\alpha}\right] \\
\overset{f}=& 1 + \frac{v_5^\alpha}{2^{\alpha R}} 
\rE_{U',V'}\tr \left[ (\cE_{5\mid U'} 
(\rho_{E \mid U',V'}))^{(1+\alpha)} \rho_{E \mid U'}^{-\alpha}\right] \\
\overset{g}\le & 1+\frac{v_5^\alpha}{2^{\alpha R}}2^{\underline{\mathrm{D}}_{1+ \alpha} \left(\rho_{UVE} \| \rho_{V-U-E} \right)},
\end{align*}
where in step $c$, $U',V'$ are distributed the same as $U,V$.
Here, $a$ follows from the fact that when $i' \neq i$ then the random variables $(V(i),V(i')) \sim p_{V(i) \mid U}\cdot p_{V(i') \mid U} $ and from the operator Jensen's inequality and from the definition of $\rho_{E \mid U}$; $b$ follows from Fact \ref{pinchs}; 
$c$ follows from the symmetry of $\cC$; 
$d$ follows because the terms in the second terms inside the trace is completely classical and from the fact that $(a+b)^x \leq a^x+b^x (a,b>0; x<1);$ $e$ follows from the circular and linear property of the trace operation; $f$ follows from the circularity of trace operation and $g$ follows from the definition of $\underline{\mathrm{D}}_{1+\alpha}(\cdot \| \cdot),$ the data-processing inequality (Fact \ref{fact:monotonequantumoperation}),  and the fact that the states involved are classical-quantum states. The desired bound now follows from the fact that $\log_2(1+x) \leq \frac{x}{\ln2}.$
\end{proof}

\section{Discussion}
We have derived an achievable rate for secure communication over fully quantum Gel'fand-Pinsker wiretap
channel. This rate is a natural quantum extension of the achievable rate of secure communication over Gel'fand-Pinsker wiretap
channel, as given in \cite{goldfled-cuff-perumter}. Here, we emphasize that 
even in the classical case, a matching converse is known only in a special case \cite[Remark 7]{goldfled-cuff-perumter}
and the question of finding a matching converse is still open in the general case.

Further, since our proof is based on an exponential upper bound of decoding error probability,
our method has the potential to improve the evaluation of error probabilities for various kinds of coding problems. Our method has two key points. The first key point is the removal of the correlations between the system $S$ and the systems $B, E$ in the analysis. 
In our method, instead of these correlations, we evaluate the correlation between the system $S$ and the message.

The second key point is composed of three types of evaluations.
The first is the evaluation of the decoding error probability of super-position coding, in which the code is randomly generated by conditional distribution.
This evaluation is based on a special type of hypothesis testing on three systems, which is discussed in Section \ref{pinchsec}.
The second is the bivariate classical-quantum channel resolvability, which is given in Section \ref{s8}.
We have evaluated the quality of approximation of the average output state when the superposition coding is applied.
Similar to the evaluation in \cite{Hayashi-wiretap}, our upper bound has the exponential from. 
The third is conditional classical-quantum channel resolvability, which is given in Section \ref{s9}.
We have derived the conditional evaluation of approximation of the average output state when the superposition coding based on $p_{UV}$ is applied.
Here, we take the condition for the choice of $V$. 
Similar to the above case, our upper bound also has the exponential form. 
Combining the first and second types of evaluations,
we have analyzed the decoding error probability of our code for our main problem.
Further, combining the second and third types of evaluations,
we have evaluated the secrecy of our code.


\bibliographystyle{ieeetr}
\bibliography{References}
\section*{Appendix}
Here we show Lemma \ref{LA} by following 7 steps.

\vspace{0.1in}

\noindent{\bf Step 1: Analysis of easy case.}\par
We first prove that 
$\max_{\rho_{UVAS} \in {\cal S}_1} R_a(\rho_{UVAS})
\ge \max_{\rho_{UVAS} \in {\cal S}_2} R_{alt}(\rho_{UVAS})$.
For $\rho_{UVAS} \in {\cal S}_2 $, 
the inequality $I[UV;B] - I[U;S]- I[V;E \mid U]\ge  I[V;B \mid U]- I[V;E \mid U]$ 
follows from $I[UV;B] - I[U;S]- I[V;E \mid U] = I[V;B \mid U]- I[V;E \mid U]+ I [U;B]-I[U;S]$ and the assumption $I [U;B]-I[U;S] \geq 0.$
Hence, 
$R_a(\rho_{UVAS})= R_{{alt}}(\rho_{UVAS})$.
Thus,
$\max_{\rho_{UVAS} \in {\cal S}_1} R_a(\rho_{UVAS})
\ge \max_{\rho_{UVAS} \in {\cal S}_2} R_a(\rho_{UVAS})
= \max_{\rho_{UVAS} \in {\cal S}_2} R_{{alt}}(\rho_{UVAS})$.

We now prove the opposite inequality
\begin{align}
\max_{\rho_{UVAS} \in {\cal S}_2} R_{{alt}}(\rho_{UVAS})
\ge 
\max_{\rho_{UVAS} \in {\cal S}_1} R_a(\rho_{UVAS}).
\label{LBF}
\end{align}
Let $\rho^{\star}_{UVAS}$ be a state such that $R_a(\rho^{\star}_{UVAS}) = \max_{\rho_{UVAS} \in {\cal S}_1} R_a(\rho_{UVAS}) >0,$ (if $R_a \leq 0$ then the bound is trivial to prove). 
If the state $\rho^{\star}_{UVAS}$ satisfies the inequality $I [U;B]-I[U;S] > 0$,
we have $
\max_{\rho_{UVAS} \in {\cal S}_1} 
R_a(\rho_{UVAS})
=R_a(\rho^\star_{UVAS})
\leq R_{{alt}}(\rho^\star_{UVAS}) \leq 
\max_{\rho_{UVAS} \in {\cal S}_2} R_{{alt}}(\rho_{UVAS}).$ 
Thus, the inequality of interest holds. 

\vspace{0.1in}

\noindent{\bf Step 2: Main part.}\par
We show \eqref{LBF} when 
$\rho^{\star}_{UVAS}$ satisfies the opposite inequality, i.e.,
\beq
\label{opposite condition}
 I [U;B]-I[U;S] \le 0. 
\enq
For this aim, we introduce the variable $\tilde{V}$ as the output of the erasure channel
of erasure probability $\eps\in [0,1]$ with the input variable $V$.
Using 
the transition probability $p_{\tilde{V} \mid V}$ of the erasure channel with erasure probability $\eps$,
we define the variables
$U' := (U, \tilde{V})$, $V' := V$, and
the following classical-quantum state: 
$$\rho_{UV\tilde{V}U'V'AS}:=\sum_{(u,v)}p_{UV}(u,v)p_{\tilde{V} \mid V}(\tilde{v} \mid v) \delta _{\left\{U' = (U,\tilde{V}), V' = V \right\}} \ket{u}\bra{u} _U\otimes \ket{v}\bra{v}_V \otimes \ket{u'}\bra{u'}_{U'} \otimes \ket{v'}\bra{v'}_{V'}\otimes \ket{\tilde{v}}\bra{\tilde{v}}_{\tilde{V}}  \otimes \rho^\star_{AS \mid u,v}.$$ 
In what follows all the calculations would be with respect to the state $\rho_{UV\tilde{V}U'V'AS}$ and the channel $\cN_{AS \to BE}.$ 
Hence, the state $ \tr_{UV\tilde{V}}[\rho_{UV\tilde{V}U'V'AS}]$ is a valid state for $R_{{alt}}.$ 
As shown later (Step 3), 
there exists an $\eps \in (0,1)$ such that 
\begin{align}
I [U';B]-I[U';S] &= 0 \label{MM1} \\
I [U;B]-I[U;S] + (1-\eps) \left( I [V;B \mid U]-I[V;S \mid U] \right)&=0.
\label{MM2}
\end{align}
Then, we choose such an $\eps \in [0,1]$.
Also, as shown later (Steps 4 and 5), we have
\begin{align}
I [U'V';B]-I[U'V';S] &\ge  R_{a}(\rho^\star_{UVAS}) \label{ab} \\
I [V';B \mid U']-I[V';E \mid U'] 
&\ge R_a(\rho^{\star}_{UVAS}) \label{ab2}.
\end{align} 
Hence, considering the definition of $R_{{alt}}(\tr_{UV\tilde{V}}[\rho_{UV\tilde{V}U'V'AS}]) $,
we obtain \eqref{LBF} as 
\begin{align*}
& \max_{\rho_{UVAS} \in {\cal S}_2} R_{{alt}}(\rho_{UVAS})
\ge 
R_{{alt}}(\tr_{UV\tilde{V}}[\rho_{UV\tilde{V}U'V'AS}]) \\
&\ge R_a(\rho^{\star}_{UVAS}) 
=
\max_{\rho_{UVAS} \in {\cal S}_1} R_a(\rho_{UVAS}).
\end{align*} 

\vspace{0.1in}

\noindent{\bf Step 3: Existence of $\varepsilon$ satisfying \eqref{MM1} and \eqref{MM2}.}\par
Now, we show that there exists an $\eps \in [0,1]$ 
satisfying \eqref{MM1} and \eqref{MM2}.
We find 
\begin{align}
I [U';B]-I[U';S] &= I [U\tilde{V};B]-I[U \tilde{V};S] \nonumber\\
&= I [U;B]-I[U;S] + I [\tilde{V};B \mid U]-I[\tilde{V};S \mid U]  \nonumber\\
\label{rrr}
&= I [U;B]-I[U;S] + (1-\eps) \left( I [V;B \mid U]-I[V;S \mid U] \right).
\end{align}
Notice that in the above if $\eps = 1$ then $I [U';B]-I[U';S] <0,$ this follows from the assumption that $I [U;B]-I[U;S] <0.$ On the other hand if $\eps = 0$ then $I [U';B]-I[U';S] = I [UV;B]-I[UV;S].$  
Our assumption that $ R_a(\rho^{\star}_{UVAS}) >0$ guarantees that $I [UV;B]-I[UV;S]>0.$ 
Thus, $I [U';B]-I[U';S] >0.$ 
Hence, we can choose an $\eps \in [0,1]$ such that $I [U';B]-I[U';S]= 0$, i.e.,
\eqref{MM1}, which implies \eqref{MM2}.

\vspace{0.1in}

\noindent{\bf Step 4: Proof of \eqref{ab}.}\par
Since $\tilde{V}$ is obtained by passing $V$ through an erasure channel, we have 
$I [UV\tilde{V};B]= I [UV;B]$
and
$I [UV\tilde{V};S]= I [UV;S]$.
Hence, we can show \eqref{ab} as follows:
\begin{align}
I [U'V';B]-I[U'V';S] &= I [UV\tilde{V};B]-I[UV \tilde{V};S] \nonumber\\
&= I [UV;B]-I[UV;S] \nonumber\\
&\overset{a} \geq 
 R_{a}(\rho^\star_{UVAS}),
\end{align} 
where $a$ follows from the definition of $R_a(\rho^\star_{UVAS})$.

\vspace{0.1in}

\noindent{\bf Step 5: Proof of \eqref{ab2}.}\par
Since $\tilde{V}$ is the output of the erasure channel of erasure probability $\eps$ 
with the input random variable $V$, 
we have
$I[\tilde{V};B \mid U]= (1-\eps) I[V;B \mid U]$
and
$I[\tilde{V};E \mid U]= (1-\eps) I[V;E \mid U]$.
Hence,
we can show \eqref{ab2} as follows:
\begin{align} 
I [V';B \mid U']-I[V';E \mid U'] & = I [V';B \mid U\tilde{V}]-I[V';E \mid U \tilde{V}] \nonumber\\
&\overset{a} = I[V;B \mid U] - I [V;E \mid U] - \left(I[\tilde{V};B \mid U] - I [\tilde{V};E \mid U]\right)\nonumber\\
& = I[V;B \mid U] - I [V;E \mid U] - (1-\eps)\left(I[V;B \mid U] - I [V;E \mid U]\right) \nonumber\\
\label{inter44}
&= \eps \left(I[V;B \mid U] - I [V;E \mid U]\right), 
\end{align}
where $a$ follows from the chain rule of mutual information. 
Using exactly similar steps we can prove that 
\beq
\label{changeEtoS}
I [V';B \mid U']-I[V';S \mid U'] = \eps \left(I[V;B \mid U] - I [V;S \mid U]\right).
\enq
As shown later (Steps 6 and 7), 
using \eqref{changeEtoS},
we have
\beq
\eps \left(I[V;B \mid U] - I [V;E \mid U]\right)
\ge R_{a}(\rho^\star_{UVAS}).
\label{changeEtoS2}
\enq
Hence, we obtain \eqref{ab2}.

\vspace{0.1in}

\noindent{\bf Step 6: Proof of \eqref{changeEtoS2} in First case.}\par
We show \eqref{changeEtoS2} when
\beq
\label{condv}
I [V;S \mid U] \geq I [V;E \mid U].
\enq
We have the following set of inequalities: 
\begin{align}
\eps \left(I[V;B \mid U] - I [V;E \mid U]\right)  
& \overset{a} \geq  \eps \left(I[V;B \mid U] - I [V;S \mid U]\right)  \nonumber\\
&\overset{b}= I [V';B \mid U']-I[V';S \mid U']  \nonumber\\
&\overset{c} = I [U' V';B]-I[U'V';S] \nonumber\\
&\overset{d}\ge 
 R_{a}(\rho^\star_{UVAS}),
\label{b1111}
\end{align}
where 
$a$ follows from \eqref{condv}; 
$b$ follows from \eqref{changeEtoS}; 
$c$ follows from \eqref{MM1}
and $d$ follows from \eqref{ab}.
Hence, we have \eqref{changeEtoS2}.

\vspace{0.1in}

\noindent{\bf Step 7: Proof of \eqref{changeEtoS2} in Second case.}\par
We show \eqref{changeEtoS2} when
\beq
\label{oppv}
I [V;S \mid U] < I [V;E \mid U].
\enq
The assumption $I[U;B] -I[U;S] \le 0$ (see \eqref{opposite condition}) implies that
\begin{align}
 I [V;B \mid U]- I[V;E \mid U] + I[U;B] -I[U;S] 
\le I [V;B \mid U]- I[V;E \mid U] \label{M51}.
\end{align}
The assumption 
$I [V;S \mid U] < I [V;E \mid U]$ (see \eqref{oppv})
implies that
\begin{align}
I[UV;B] - I[U;S]- I[V;E \mid U] 
<  I[UV;B] - I[UV;S] \label{M52}.
\end{align}
Using the chain rule, we have 
\begin{align}
I[UV;B] - I[U;S]- I[V;E \mid U] = I [V;B \mid U]- I[V;E \mid U] + I[U;B] -I[U;S] .
\label{M53}
\end{align}
Combining \eqref{M51}, \eqref{M52}, and \eqref{M53},
we have
\begin{align}
I[UV;B] - I[U;S]- I[V;E \mid U] 
\le \min\{I [V;B \mid U]- I[V;E \mid U], I[UV;B] - I[UV;S]\}
\label{M54}.
\end{align}
Hence, the minimum $R_a(\rho^{\star}_{UVAS})$ is given as
\begin{align}
R_a(\rho^{\star}_{UVAS})
=I[UV;B] - I[U;S]- I[V;E \mid U] 
\label{M55}.
\end{align}

Now, we have the following set of inequalities: 
\begin{align}
\eps \left(I[V;B \mid U] - I [V;E \mid U]\right) 
&= I[V;B \mid U] - I [V;E \mid U] - (1-\eps) \left(I[V;B \mid U] - I [V;E \mid U]\right) \nonumber\\
&\overset{a}> I[V;B \mid U] - I [V;E \mid U] - (1-\eps) \left(I[V;B \mid U] - I [V;S \mid U]\right) \nonumber\\
&\overset{b} = I[V;B \mid U] - I [V;E \mid U] + I(U;B) - I(U;S) \nonumber\\
& \overset{c} = 
R_a(\rho^{\star}_{UVAS})
\label{b2222},
\end{align}
where
$a$ follows from the assumption that $I [V;S \mid U] < I [V;E \mid U]$ (see \eqref{oppv}); 
$b$ follows from \eqref{MM2}
and 
$c$ follows from \eqref{M55}.
Hence, we obtain \eqref{changeEtoS2}.
This completes the proof. 

\end{document}